\documentclass[journal]{IEEEtran}
\usepackage{diagbox}
\usepackage{bm}
\usepackage{cite}

\ifCLASSINFOpdf
  \usepackage[pdftex]{graphicx}
  \graphicspath{{../pdf/}{../jpeg/}}
  \DeclareGraphicsExtensions{.pdf,.jpeg,.png}
\else
  \usepackage[dvips]{graphicx}
  \graphicspath{{../eps/}}
  \DeclareGraphicsExtensions{.eps}
\fi
\usepackage{amsmath}
\usepackage{array}

\usepackage{tabularx,booktabs}
\usepackage{multirow}

\makeatletter
\let\NAT@parse\undefined
\makeatother
\usepackage{hyperref}  

\begin{document}
%
\title{Coverless Video Steganography based on Maximum DC Coefficients}
%
%
%

\author{Laijin~Meng,~
        Xinghao~Jiang,~\IEEEmembership{Senior~Member,~IEEE}
        Zhenzhen~Zhang,~
        Zhaohong~Li,~
        and~Tanfeng~Sun,~\IEEEmembership{Senior~Member,~IEEE}
\thanks{This work was supported in part by the Scientific Research Common Program of Beijing Municipal Commission of Education (No. KM202110015004), National Natural Science Foundation of China (No.61771270, No.61572320, No.61702034), the National Key Research and Development Projects of China under Grant 2018YFC0830700 and Zhejiang Provincial Natural Science Foundation of China (LR20F020001) .}
\thanks{Laijin Meng, Xinghao Jiang, and Tanfeng Sun are with National Engineering Lab on Information Content Analysis Techniques, School of Electronic Information and Electrical Engineering, Shanghai Jiao Tong University, Shanghai, China. (e-mail: menglaijin@sjtu.edu.cn, xhjiang@sjtu.edu.cn, tfsun@sjtu.edu.cn).}
\thanks{Zhenzhen Zhang is with School of Information Engineering, Beijing Institute of Graphic Communication, Beijing, China. (e-mail: zhangzhenzhen@bigc.edu.cn).}
\thanks{Zhaohong Li is with School of Electronic and Information Engineering, Beijing Jiaotong University, Beijing, China. (e-mail: zhhli2@bjtu.edu.cn).}
\thanks{Corresponding authors are Xinghao Jiang and Zhenzhen Zhang.}}

%
%

\markboth{Journal of \LaTeX\ Class Files,~Vol.~14, No.~8, August~2015}%
{Shell \MakeLowercase{\textit{et al.}}: Bare Demo of IEEEtran.cls for IEEE Journals}
%



\maketitle

\begin{abstract}
Coverless steganography has been a great interest in recent years, since it is a technology that can absolutely resist the detection of steganalysis by not modifying the carriers. However, most existing coverless steganography algorithms select images as carriers, and few studies are reported on coverless video steganography. In fact, video is a securer and more informative carrier. In this paper, a novel coverless video steganography algorithm based on maximum Direct Current (DC) coefficients is proposed. Firstly, a Gaussian distribution model of DC coefficients considering video coding process is built, which indicates that the distribution of changes for maximum DC coefficients in a block is more stable than the adjacent DC coefficients. Then, a novel hash sequence generation method based on the maximum DC coefficients is proposed. After that, the video index structure is established to speed up the efficiency of searching videos. In the process of information hiding, the secret information is converted into binary segments, and the video whose hash sequence equals to secret information segment is selected as the carrier according to the video index structure. Finally, all of the selected videos and auxiliary information are sent to the receiver. Especially, the subjective security of video carriers, the cost of auxiliary information and the robustness to video compression are considered for the first time in this paper. Experimental results and analysis show that the proposed algorithm performs better in terms of capacity, robustness, and security, compared with the state-of-the-art coverless steganography algorithms.
\end{abstract}


%
\IEEEpeerreviewmaketitle

\section{Introduction}
%
%
%
%
\IEEEPARstart{I}{nformation} security is increasingly drawing people’s attention in recent years due to the globalization of information and communication technology. Cryptography is the very first exploited means to keep the security of media information. However, it is prone to be perceived and further deciphered because of its scrambling characteristic. To better protect the security of information data, steganography, the technology of hiding secret information into the host media such as audio, digital images, and videos without raising suspicion, has been developed and is becoming a hot issue in the area of information security.
\par Traditional steganography \cite{1,2,3,4,5,6,7,8,9,10,11,12} embeds secret information by modifying carriers’ parameter characteristics. Usually, these modifications are too minor to be distinguished by human eyes, but can be detected by some special steganalysis algorithms \cite{13,14,15,16,17} or universal steganalysis algorithms \cite{18,19}. In order to solve this problem completely, the concept of coverless steganography was proposed. It does not mean that the process of embedding secret information can be accomplished without carriers. In fact, it establishes mapping relationship between secret information and carriers. Since the carrier has not been modified in coverless steganography, it can avoid being detected by steganalysis algorithms fundamentally \cite{20}, which makes coverless steganography become a securer way for data hiding compared with traditional steganography.
\par The existing coverless steganography schemes mainly used images as carriers. Zhou et al. \cite{20} designed the first coverless image steganography, which divided the image into 3$\times$3 non-overlapping blocks and generated hash sequences by comparing average intensity of adjacent blocks. After that, a number of coverless image steganography schemes are proposed. Zheng et al. \cite{21} introduced Scale-Invariant Feature Transform (SIFT) features in the process of generating hash sequences, while Yuan et al. \cite{22} utilized SIFT features and bag-of-features at the same time. Besides, Zhou et al. \cite{23} used a set of proper partial duplicates of the given secret image as stego-images. To further improve the performance of \cite{23}, Luo et al. \cite{24} proposed a novel coverless steganography based on image block matching and dense convolutional network. Recently, the robustness and security are of greater concern, so Zhang et al. \cite{25} proposed a robust coverless image steganography algorithm based on discrete cosine transform (DCT) and latent Dirichlet allocation (LDA) topic classification. Then Liu et al. \cite{26} proposed a coverless steganography algorithm based on image retrieval of DenseNet features and DWT sequence mapping to improve the robustness of \cite{25}.
\par Compared with images, video sequences not only contain spatial information, but also have abundant temporary redundancy. However, few studies of coverless video steganography have been reported. As far as we know, until 2020, Pan et al. \cite{27} proposed the first coverless video steganography method. In this work, they calculated the statistical histogram of semantic information by semantic segmentation neural network. Then, hash sequences were generated by comparing semantic proportion of the histogram. Moreover, they defined the criterion of effective capacity for the first time, but the experimental results show that there was a certain gap between the capacity of this work and the theoretical capacity. Besides low capacity, the robustness to some image attacks and some common operations or attacks strongly related to video, such as video compression and frame deletion, were not considered in the only existing coverless video steganography \cite{27}. Furthermore, most of the existing coverless image steganography schemes \cite{25,26} can resist image noise and filtering attacks, but they cannot resist video compression effectively when their hash sequence generation methods are used directly for video sequences. So, there is no existing way that can be used in coverless video steganography for resisting video compression and other video attacks.
\par To solve this problem, we first analyze the reconstruction error in video compression and build a Gaussian distribution model to fit the change rates of Direct Current (DC) coefficients. From the model, we find that the distribution of changes for maximum DC coefficients in a block is more concentrated than the adjacent DC coefficients under the video compression attack, which indicates that it may be more robust to use maximum DC coefficients to generate hash sequences in coverless steganography. According to the theoretical analysis, a robust coverless video steganography algorithm based on maximum DC coefficients is proposed in this paper. 
\par The video frame is partitioned into non-overlapping blocks firstly. Then each block is further partitioned into 4$\times$4 sub-blocks. After that, hash sequences are generated by comparing each DC coefficient with maximum DC coefficient in a sub-block with raster scan order. Finally, an index structure for videos is established. Besides video compression, frame deletion attack is also considered in this paper as it is a popular operation to video sequences. Experimental results show that the proposed steganography can achieve better resistance against most image attacks such as rotation, scaling, and noise compared with the state-of-the-art work of both image and video coverless steganography methods \cite{25,26,27}. It is worth mentioning that the proposed algorithm performs much better in resisting video compression and frame deletion attack. Moreover, the proposed method achieves both larger absolute and relative effective capacity, and the latter one is a fairer evaluation metric which is proposed for the first time in this paper.
\par As we know, security is the first important thing for coverless steganography. On the one hand, coverless steganography cannot be detected by steganalysis algorithms, and on the other hand, sending a great quantity of images can be suspicious \cite{20,21,22,23,24}, even though they are similar in content \cite{25,26}. In order to clarify this issue, we tested the subjective security by giving out questionnaires. The experimental results show that sending a video has more advantages compared with sending a group of images.
\par In general, the main contributions of this paper can be concluded as follows.
\par 1) It is the first time to test the subjective security of coverless steganography, not limited to the objective security test of its ability to resist steganalysis. The results show that the video has more advantages than images. However, video-based coverless steganography has received little attention. As far as we know, there was no relevant report until June 2020, but this work did not elaborate and investigate the subjective security of video carriers, which is very important for certifying the security of coverless steganography with videos. Based on the investigation in this paper, the security advantages of video will attract attention, providing a new and safer hidden transmission method for coverless steganography.
\par 2) A Gaussian distribution model of DC coefficients considering video coding process is built, which indicates that the distribution of changes for maximum DC coefficients in a block is more stable than the adjacent DC coefficients. On this basis, a new coverless video steganography algorithm based on maximum DC coefficients is proposed, which can resist common attacks very well. It is worth emphasizing that video compression and frame deletion attacks are considered for the first time and the proposed steganography can resist these two video attacks much better than the state-of-the-art work of image and video coverless steganography algorithms.
\par 3) A new practical and fair metric, relative effective capacity is proposed for evaluating the capacity of steganography algorithms in this paper, which is defined as the ratio of effective capacity and the length of auxiliary information. For the existing coverless steganography schemes \cite{20,21,22,23,24,25,26,27}, auxiliary information needs to be sent with secret information. However, if the length of auxiliary information is too long compared with the secret one, for example, 10 bits compared with 20 bits, it will lose meaning of coverless steganography. So, except the absolute secret information bits, the relative effective capacity is also important for measuring a coverless steganography algorithm. Owing to the abundance of redundancy in videos and the method of maximum DC coefficients based on block partitioning, the proposed algorithm performs better in both absolute and relative effective capacity compared with the state-of-the-art work.
\par The rest of the paper is arranged as follows. Subjective security test is shown in Section \uppercase\expandafter{\romannumeral2}. The distribution of changes for DC coefficients after video compression is modeled and a novel hash sequence generation method is proposed in Section \uppercase\expandafter{\romannumeral3}. The proposed coverless video steganography is presented in Section \uppercase\expandafter{\romannumeral4}. The experimental results and analysis are described in Section \uppercase\expandafter{\romannumeral5}. Conclusions are drawn in Section \uppercase\expandafter{\romannumeral6}.

\section{Subjective Security Test}
\par As we know, security is the most important issue to be considered for both traditional steganography and coverless steganography. As mentioned above, coverless steganography cannot be detected by steganalysis algorithms due to no modification to carriers. Besides the objective steganalysis detection, subjective security is also an important evaluation criterion, which refers to whether the way or content of transmission will arouse doubts by human eyes. In the existing coverless steganography, only few algorithms \cite{25,26} considered subjective security, but they did not do a specific experimental test of subjective security in practical.
\par In this paper, we do a subjective security test to collect people’s views on the transmission of different carriers. The test method is illustrated first, and then the result is discussed.

\subsection{Test Method}
\par Most of the existing coverless steganography algorithms are image-based, and they required to transmit a large series of images which leads to subjective security issues. In previous works, LDA topic model \cite{25} and retrieval system of Densenet convolutional neural work \cite{26} were utilized to improve the similarity between transmitted images, but these methods did not solve this problem fundamentally. Compared with images, video sequences include abundant temporary redundancy, which brings strong frame similarity. However, video-based coverless steganography has received little attention. As far as we know, there has not been a work discussing the subjective security of video carriers. In this part, we make a questionnaire to investigate the views of different people about the security of using different kinds of carriers for coverless steganography.
\par We record three videos of transmitting a series of content irrelevant images, a series of content relevant images, and a video respectively to simulate how people communicate with each other through social software (The questionnaire can be accessed at https://wj.qq.com/s2/6901439/d343). After that, we set up three options of the most suspicious, the medium suspicious, and the least suspicious for people to make choice. 

\subsection{Test Result}
\par
\begin{figure}[!t]
	\centering
	\includegraphics[width=0.5\textwidth]{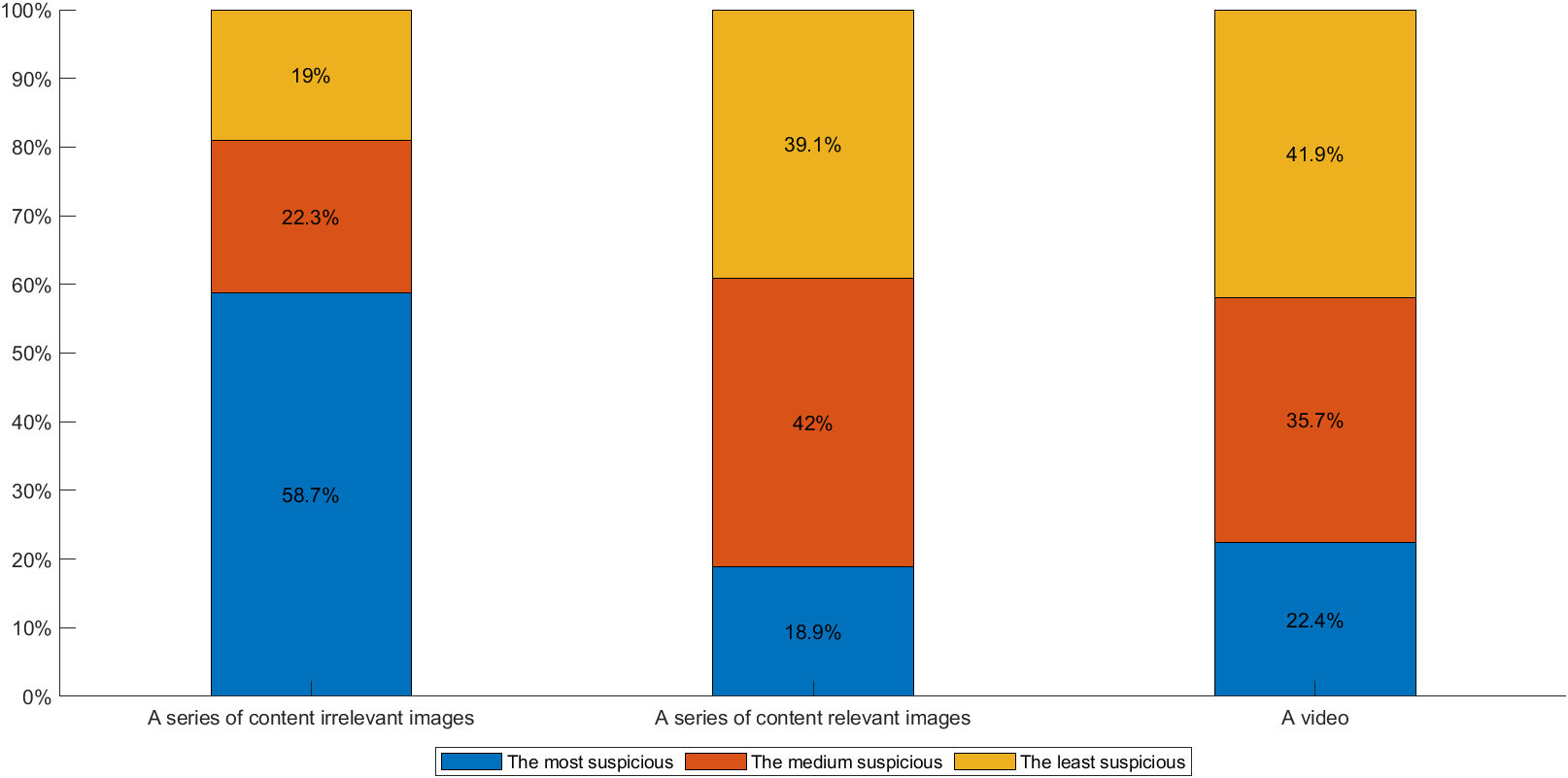}
	\caption{The proportion of three levels of suspicion for carrier selections.}
	\label{fig_1}
\end{figure}

 According to the above test method, we collected 242 valid questionnaires in total. Then we calculated the proportion of three levels of suspicion for different carrier selections. The result is shown in Fig. \ref{fig_1}.
\par It can be seen that most people (58.7\%) think that transmitting a series of content irrelevant images brings the highest suspicion. In the option of transmitting a series of content relevant images, the proportion of medium suspicious accounts for the largest, and is a little higher than that of the least suspicious, which indicates transmitting a large number of content relevant images are securer than irrelevant images, but still will arouse human's suspicion. As for a video, the least suspicious option possesses the largest proportion, which indicates that the subjective security advantage of transmitting videos as carriers for coverless steganography. Besides, from the view of constructing datasets, it is more difficult to obtain a large number of content relevant images compared with videos.
\par Based on the advantage of videos compared with images, we choose videos as carriers and develop a novel video-based coverless steganography.

\section{Distribution Model of DC Coefficients in Video Coding}
\par In order to achieve strong robustness under various attacks, it is the most important step to extract appropriate robust features to generate hash sequences for coverless steganography. DCT, in particular the DCT-\uppercase\expandafter{\romannumeral2}, is often used in many coding standards, because of its strong “energy compaction” property \cite{28}. Taking video coding as an example, DCT has been applied since H.261 and it has been retained in subsequent video coding standards. Since most of the energy is concentrated in low frequency coefficients, the changes of DC coefficient will be much slighter compared with AC coefficients under a variety of image operations, so DC coefficients were usually adopted to generate hash sequences in coverless image steganography algorithms \cite{24,25}. However, the hash sequence generation methods in these algorithms are basically based on the relationship of adjacent DC coefficients. We found that this generation method can resist most image attacks, but for the video carrier, the relationship between adjacent DC coefficients will change significantly before and after video compression, and the robustness is much weaker. So, it is necessary to study the distribution of changes for DC coefficients after video compression in order to propose a more robust hash sequence generation method.
\par For analyzing the change of DC coefficients before and after video compression, we need to find out what causes the change. Therefore, according to the general video compression process, we derive the difference of DC coefficients before and after video coding from the perspective of reconstruction error, the specific process is as follows.
\par Suppose that the original frame is \bm{$F$}, whose size is \emph{M}$\times$\emph{N}, \bm{$F'$} is the decoded frame of \bm{$F$} after video compression. In the process of video compression, coding is block-based. Consider High Efficiency Video Coding (HEVC) as an example, Coding Unit (CU) will be partitioned into non-overlap Prediction Units (PUs) and Transform Units (TUs) for prediction and transforming. For simplicity, we suppose that one CU is divided into one PU and one TU, and the derivation of other cases is similar to this one. The input pixel values of luminance component in a CU is denoted as \bm{$\!Y\!B$}, and the reconstruction component after compression \bm{$\!Y\!B'$} can be derived by
\setlength{\arraycolsep}{0.0em}
\begin{eqnarray}
\label{eqn_1}
\bm{\!Y\!B'}&=&RT\{IDCT([DCT(\bm{Y\!B}-\bm{RB})/QP]\times QP)\}+\bm{RB}\nonumber\\
&\approx& RT\{IDT([DCT(\bm{Y\!B})/QP]\times QP)\}\nonumber\\
&-&RT\{IDT([DCT(\bm{RB})/QP]\times QP)\}+\bm{RB}\nonumber\\
&=&\bm{Y\!B}+E(\bm{Y\!B})-E(\bm{RB}) ,
\end{eqnarray}
\setlength{\arraycolsep}{5pt}where \bm{$RB$} is the reference block of \bm{$Y\!B$} in the corresponding inter-prediction process, and $QP$ means the quantization step. \emph{DCT(.)} and \emph{IDCT(.)} stand for discrete cosine transform (DCT) and inverse DCT, respectively. What’s more, $[.]$ denotes the rounding operation while $RT\{.\}$ represents rounding and  truncating operations. $E(\bm{Y\!B})$ and $E(\bm{RB})$ denote the irreversible quantization error and reconstruction error of \bm{$\!Y\!B$} and \bm{$RB$}, which shows that there is some irreversible error between \bm{$\!Y\!B$} and \bm{$\!Y\!B'$}. Specifically, the greater $QP$ is, the greater the loss due to rounding and truncation operations, which makes a greater difference between \bm{$\!Y\!B$} and \bm{$\!Y\!B'$}. 
\par For the whole original frame \bm{$F$}, the error of the decoded frame \bm{$F'$} after compression is accumulated by all TUs, so the relationship between \bm{$F$} and \bm{$F'$} can be represented as
\begin{equation}
\label{eqn_2}
\bm{F'}=\bm{F}+E(\bm{F})-E(\bm{C}) ,
\end{equation}
where \bm{$C$} is the reference frame of \bm{$F$}, $E(\bm{F})$ and $E(\bm{C})$ are irreversible quantization error and reconstruction error under all TUs' accumulation, respectively.
\par In the existing hash sequence generation method of coverless image steganography, block partitioning is implemented twice firstly, and then DC coefficients are extracted. Finally, hash sequences are generated by comparing adjacent DC coefficients. The specific process can be described as follows.
\par For the original frame \bm{$F$}, it can be partitioned into \emph{m}$\times$\emph{n} blocks. The block $\bm{Y}_{b}$ can be represented with raster scan order
\begin{equation}
\label{eqn_3}
\bm{Y}_{b}=\{\bm{Y}_{1},\bm{Y}_{2},\ldots, \bm{Y}_{m\times n}\}, 0<b\le m\times n .
\end{equation}
\par For each block $\bm{Y}_{b}$, Y channel is extracted and further partitioned into 16 sub-blocks. The sub-block $\bm{y}_{i}$ can be obtained with raster scan order
\begin{equation}
\label{eqn_4}
\bm{y}_{i}=\{\bm{y}_{1},\bm{y}_{2},\ldots, \bm{y}_{16}\}, 0<i\le 16 .
\end{equation}
\par After that, 8$\times$8 DCT is done for each sub-block $\bm{y}_{i}$
\setlength{\arraycolsep}{0.0em}
\begin{eqnarray}\label{eqn_5}
&\bm{C}_{j}(u,v)=DCT\left(\bm{B} \bm{l}_{j}\right), 0<j \leq\left(\frac{M\times N}{32 m\times 32 n}\right),\nonumber \\
&0<u,v\leq 8 ,
\end{eqnarray}
\setlength{\arraycolsep}{5pt}where $\bm{Bl}_{j}$ is the \emph{j}-th 8$\times$8 block in the sub-block $\bm{y}_{i}$ with raster scan order, and $\bm{C}_{j}(u,v)$ is the corresponding DCT coefficient. Then $\bm{DC}_{i}$ is obtained by adding the DC coefficients of all 8$\times$8 blocks in $\bm{y}_{i}$
\begin{equation}
\label{eqn_6}
\bm{DC}_{i}=\sum_{j=1}^{\left(\frac{M \times N}{32 m \times 32 n}\right)} \bm{C}_{j}(0,0) .
\end{equation}
\par According to $\bm{DC}_{i}$ of the adjacent sub-blocks, each bit of hash sequence $\bm{\!H\!S}_{b}$ of each block $\bm{Y}_{b}$ can be achieved as
\begin{equation}
\label{eqn_7}
\bm{\!H\!S}_{b}^{i}=\left\{\begin{array}{ll}
1, & \text { if } \bm{DC}_{i}>\bm{DC}_{i+1} \\
0, & \text { if } \bm{DC}_{i} \leq \bm{DC}_{i+1}
\end{array}, 0<i<16. \right.
\end{equation}
\par To obtain the DC coefficients of decoded frame \bm{$F'$} after video compression, the same operation of frame \bm{$F$}, twice block partitioning is made for decoded frame \bm{$F'$}. Let $\bm{Bl'}_{j}$ stand for the block in corresponding position of \bm{$F'$}, as derived from (\ref{eqn_1}), the following relationship between $\bm{Bl'}_{j}$ and $\bm{Bl}_{j}$ can be achieved
\begin{equation}
\label{eqn_8}
\boldsymbol{Bl}_{j}^{\prime}=\boldsymbol{B} \boldsymbol{l}_{j}+E\left(\boldsymbol{B} \boldsymbol{l}_{j}\right)-E\left(\boldsymbol{C\!B}_{j}\right) ,
\end{equation}
where $\bm{\!C\!B}_{j}$ is the reference block of $\bm{Bl}_{j}$ in the reference frame $\bm{C}$, and as in (\ref{eqn_1}), here, $E(\bm{Bl}_{j})$ and $E(\bm{C\!B}_{j})$ denote the irreversible quantization error and reconstruction error of blocks $\bm{Bl}_{j}$ and $\bm{\!C\!B}_{j}$. Note that for I pictures, there is no inter prediction mode, so the prediction error of $E(\bm{C\!B}_{j})$ equals to 0. 

\par Then DCT coefficients of decoded block $\bm{Bl'}_{j}$ can be calculated as
\setlength{\arraycolsep}{0.0em}
\begin{eqnarray}\label{eqn_9}
\bm{C}_{j}^{\prime}(u, v)&=&DCT(\bm{Bl}_{j}^{\prime}) \nonumber\\
&=&DCT(\bm{B l}_{j}+E(\bm{Bl}_{j})-E(\bm{C\!B}_{j})) \nonumber\\
&\approx& DCT(\bm{Bl}_{j})+DCT(E(\boldsymbol{Bl}_{j}))-DCT(E(\boldsymbol{C\!B}_{j})), \nonumber\\
&0&<j \leq\left(\frac{M \times N}{32 m \times 32 n}\right), 0<u, v \leq 8 .
\end{eqnarray}
\setlength{\arraycolsep}{5pt}
\par
\begin{figure}[!t]
	\centering
	\includegraphics[width=0.5\textwidth]{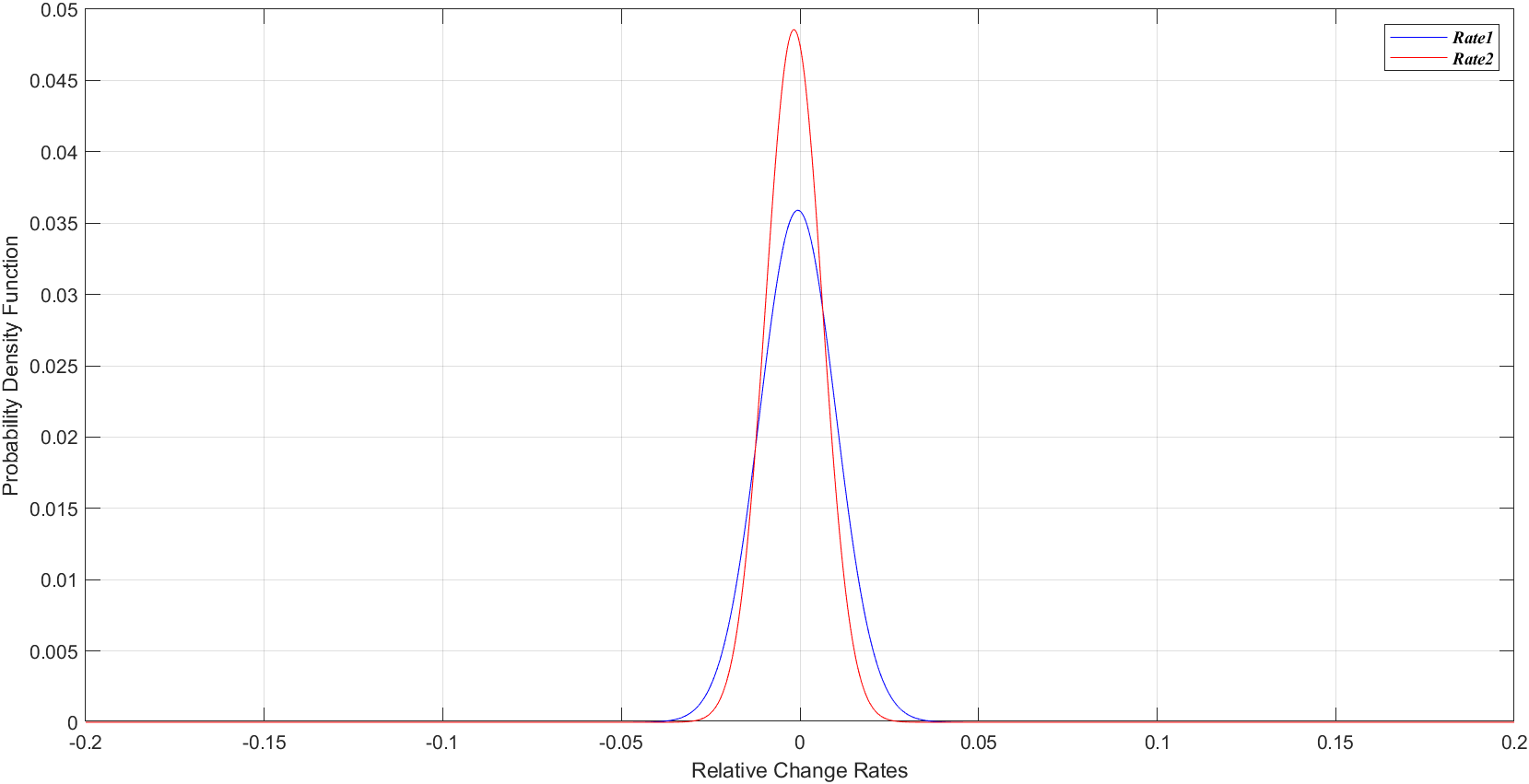}
	\caption{Fitted probability density function of $\bm{Rate1}$ and $\bm{Rate2}$.}
	\label{fig_2}
\end{figure}

 After that, the sum of DC coefficients, $\bm{DC'}_{i}$ can be achieved as
\setlength{\arraycolsep}{0.0em}
\begin{eqnarray}\label{eqn_10}
\bm{DC}_{i}^{\prime}&=&\sum_{j=1}^{\left(\frac{M \times N}{32 m \times 32 n}\right)} \bm{C}_{j}^{\prime}(0,0) \nonumber\\
&=&\sum_{j=1}^{\left(\frac{M \times N}{32 m \times 32 n}\right)} \bm{C}_{j}(0,0)+DC\left(DCT\left(E\left(\bm{B l}_{j}\right)\right)\right)\nonumber\\
&-&DC\left(DCT\left(E\left(\boldsymbol{C\!B}_{j}\right)\right)\right),
\end{eqnarray}
\setlength{\arraycolsep}{5pt}where \emph{DC(.)} denotes the function of obtaining the DC coefficient in one DCT coefficient matrix. From the above analysis, we can see that the change of DC coefficients after video compression is caused by the quantization error and reconstruction error. To further analyze the changes of DC coefficients under video compression, we define change rates of DC coefficients, and then a Gaussian distribution model is utilized to fit change rates of DC coefficients, which can be derived as follows
\setlength{\arraycolsep}{0.0em}
\begin{eqnarray}\label{eqn_11}
&\bm{Rate1}=\frac{\bm{DC}_{i}^{\prime}-\bm{DC}_{i}}{\bm{DC}_{i}} \nonumber\\
&=\sum_{j=1}^{\left(\frac{M \times N}{32 m \times 32 n}\right)} \frac{DC\left(DCT\left(E\left(\boldsymbol{Bl}_{j}\right)\right)\right)-DC\left(DCT\left(E\left(\boldsymbol{C\!B}_{j}\right)\right)\right)}{\boldsymbol{C}_{j}(0,0)} ,
\end{eqnarray}
\setlength{\arraycolsep}{5pt}
\setlength{\arraycolsep}{0.0em}
\begin{eqnarray}\label{eqn_12}
&\bm{Rate2}=\frac{\boldsymbol{D} \boldsymbol{C}_{\max }^{\prime}-\boldsymbol{D} \boldsymbol{C}_{\max }}{\boldsymbol{D} \boldsymbol{C}_{\max }} \nonumber\\
&=\frac{\max \left\{DC_{1}{ }^{\prime}, DC_{2}{ }^{\prime}, \ldots, DC_{16}{ }^{\prime}\right\}-\max \left\{DC_{1}, DC_{2}, \ldots, DC_{16}\right\}}{\max \left\{DC_{1}, DC_{2}, \ldots, DC_{16}\right\}} ,
\end{eqnarray}
\setlength{\arraycolsep}{5pt}where $\bm{DC}_{max}$ and $\bm{DC'}_{max}$stand for the maximum of $\bm{DC}_{i}$ and $\bm{DC'}_{i}$, respectively. Thus, $\bm{Rate1}$ represents the change rate of every DC coefficient in one block, and $\bm{Rate2}$ represents the change rate of the maximum DC coefficient of the block. To illustrate it more vividly, the distribution of $\bm{Rate1}$ and $\bm{Rate2}$ with $QP=37$ , and $m = 26$, $n=5$  for videos of Class C dataset (The specific videos of Class C will be described in experimental results.) is calculated and modelled by the Gaussian distribution as shown in Fig. \ref{fig_2}. The general form of its probability density function (PDF) is
\begin{equation}
\label{eqn_13}
f(x)=\frac{1}{\sigma \sqrt{2 \pi}} \exp \left(-\frac{1}{2}\left(\frac{x-\mu}{\sigma}\right)^{2}\right) ,
\end{equation}
where the parameter $\mu$ and $\sigma$ are the mean and standard deviation of the distribution, respectively. 

\par We can see that compared with $\bm{Rate1}$, the distribution of $\bm{Rate2}$ is more concentrated in Fig. \ref{fig_2}. More specifically, the standard deviations of $\bm{Rate1}$ and $\bm{Rate2}$ equal to 0.015 and 0.011, respectively. Numerically, a smaller standard deviation also means a more concentrated distribution. From the analysis of video coding from (\ref{eqn_1}) to (\ref{eqn_12}), we know the change of DC coefficients is caused by irreversible errors, so it is impossible to avoid the changes. But from the distribution model, it can be seen that the maximum DC coefficient performs more stable after video compression, which indicates using maximum DC coefficients to generate hash sequence can improve the robustness to video compression compared with the traditional way of using the adjacent DC coefficients in most existing coverless image steganography algorithms. Based on this suppose, a novel hash sequence generation method is proposed
\begin{equation}
\label{eqn_14}
\bm{\!H\!S}_{b}^{i}=\left\{\begin{array}{l}
1, if \frac{\bm{DC}_{i}}{\bm{DC}_{\max }}>T \vspace{1ex}\\
0, if \frac{\bm{DC}_{i}}{\bm{DC}_{\max }} \leq T
\end{array}, 0<i<16 ,\right.
\end{equation}
where \emph{T} is the threshold, which is set to meet the ratio of bits 0 and 1 in hash sequence close to 1:1, and it will be discussed in experimental results.
\par In short, in this section, some preliminaries about DCT and DC coefficients are introduced firstly. Then the reconstruction error in video compression is analyzed, and on this basis a Gaussian distribution model is utilized to prove the robustness of maximum DC coefficients. Finally, a novel hash sequence generation method based on maximum DC coefficients is proposed.
\begin{figure}[!t]
	\centering
	\includegraphics[width=0.5\textwidth]{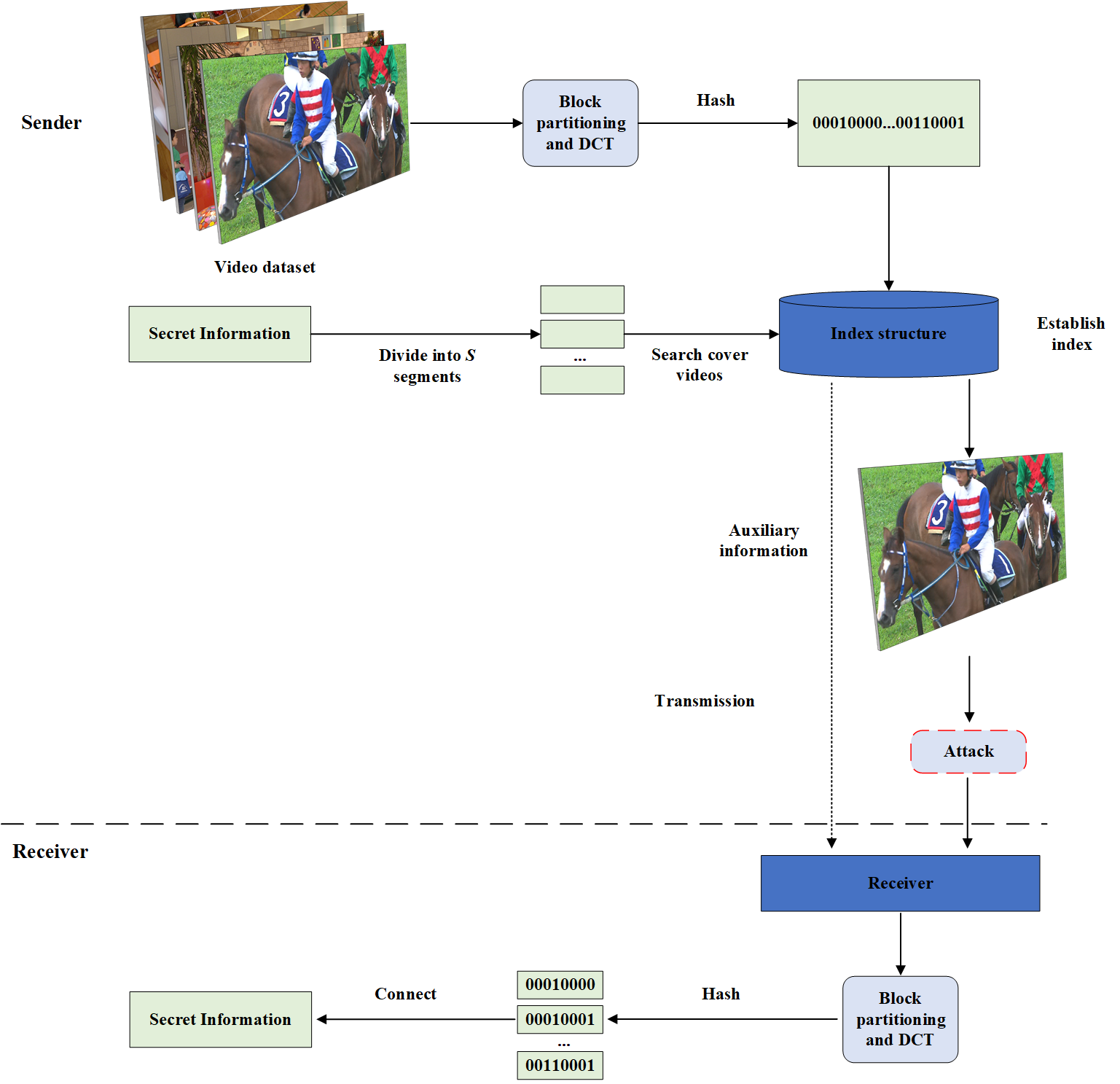}
	\caption{The framework of the proposed coverless video steganography.}
	\label{fig_3}
\end{figure}
\section{The Proposed Coverless Video Steganography}
\subsection{Framework of Coverless Video Steganography}
\par The proposed coverless video steganography is composed of four main modules as shown in Fig. \ref{fig_3}, which are the generation of hash sequences based on DCT, the establishment of video index structure, the secret information hiding and the extraction of secret information, respectively.
\par For the sender, block partitioning is implemented twice firstly. After that, 8$\times$8 DCT is performed for each Y channel of sub-blocks to extract DC coefficients as illustrated in Section \uppercase\expandafter{\romannumeral3} from (\ref{eqn_3})-(\ref{eqn_6}). And then, hash sequences like 00010000…00110001 can be achieved with the proposed hash sequence generation method which compares DC coefficients with the maximum DC coefficient as in (\ref{eqn_14}). Meanwhile, the secret information is divided into \emph{S} segments with the length of \emph{L}. For each segment, the corresponding video whose hash sequence is the same as the secret information segment will be chosen as the carrier by searching the established video index database, which will be illustrated later. Finally, all of the selected videos and corresponding auxiliary information which includes location information and video ID will be sent to the receiver. Noting that the total length of the sequence may not be a multiple of \emph{L}, 0 will be padded for the last segment to form a sequence of length \emph{L}, and the number of 0 will be recorded as a special part of auxiliary information.
\par As for the receiver, the corresponding blocks, frames, and videos are chosen according to the auxiliary information firstly, and then the proposed feature extraction method and hash sequence generation method are implemented to obtain the segmented secret information sequences. After that, the segmented sequences are connected to recover the original secret information. 
\subsection{Establishment of Video Index Database}
\par Hash sequences are a series of fixed-length sequences extracted from the carriers. Furthermore, hash sequence generation method is the most important part in coverless steganography, which directly determines the capacity and robustness of the steganography algorithm. As described in Section \uppercase\expandafter{\romannumeral3}, a novel hash sequence generation method based on maximum DC coefficients is proposed. According to (\ref{eqn_3})-(\ref{eqn_6}) and (\ref{eqn_14}), the hash sequences of all blocks for each frame in video dataset can be obtained.
\par The principle of coverless steganography is sending the carrier whose hash sequence equals to the secret information segments. Usually, the secret information has to be divided into lots of segments, and as a matter of fact, it will cost a lot of time to find the video carriers for all the secret segments, especially when the video database is huge. So, the video index database is built to speed up the efficiency of searching videos. The video index database is an artificial structure which is similar to the index structure of dictionary. The corresponding video can be found by indexing the hash sequence. In fact, it is realized by the tree structure, which is a common but efficient data structure.  
\par
\begin{figure}[!t]
	\centering
	\includegraphics[width=0.5\textwidth]{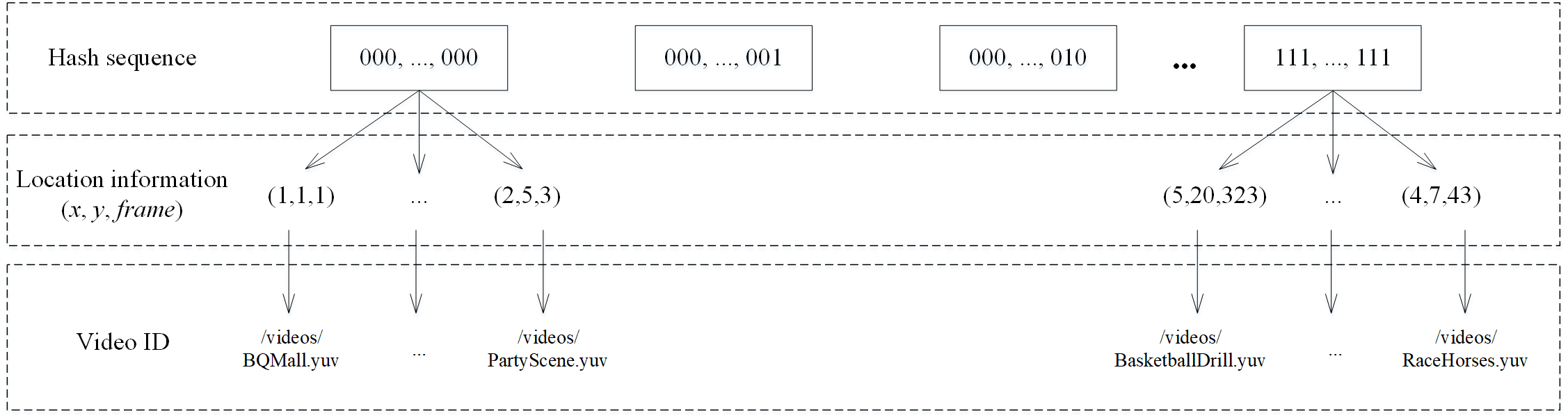}
	\caption{The diagram of the video index database.}
	\label{fig_4}
\end{figure}

 The established video index database is shown in Fig. \ref{fig_4}. The whole structure can be divided into three parts, which are hash sequence, location information, and video ID. Hash sequence is the entrance of the video index database, which ranges from “000, …, 000” to “111, …, 111” in binary incremental order. Location information contains three coordinates. The first two coordinates \emph{x} and \emph{y} indicate the position of the image block $\bm{Y}_{b}$ in one video frame, while the latter one \emph{frame} represents the corresponding frame order in a video sequence. The last part video ID stores the path to the folder and the video file name. To illustrate it more clearly, we assume that the hash sequence of blocks at the (1, 1) coordinates in the first frame of video BQMall is “000, …, 000”, and then the leftmost branch in Fig. \ref{fig_4} is established. The similar process is repeated until all videos are listed in the video index database. 
\subsection{Secret Information Hiding}
\par In coverless video steganography, the process of secret information hiding is the procedure of selecting corresponding videos as carriers based on the relationship between secret information and hash sequences. In this part, the specific process of information hiding is introduced.
\par \emph{Step} 1. Suppose that the total length of the secret information \bm{$B$} is \emph{h}, the secret information is divided into \emph{S} binary information segments
\begin{equation}
\label{eqn_15}
S=\left\{\begin{array}{ll}
h/L, \quad &if~h \% L=0 \vspace{1ex}\\
\lfloor h/L\rfloor\!+\!1, \quad &Otherwise
\end{array} ,\right.
\end{equation}
where \emph{L} is the length of the information segment and $\lfloor .\rfloor$ represents a downward rounding function. When the total length \emph{h} cannot be divided by \emph{L}, zeros will be padded for the last segment to form a sequence of length \emph{L}, and the number of zeros is recorded.
\par \emph{Step} 2. The hash sequence of each video is obtained according to the proposed hash sequence generation method in Section \uppercase\expandafter{\romannumeral3}. Then, the video index database can be established as described in Section \uppercase\expandafter{\romannumeral4}.\emph{C}. After that, for a secret information segment $\bm{I}_{s}, s=1,2,\ldots,S$, the video which has the image block whose hash sequence is the same as $\bm{I}_{s}$ is selected, and the auxiliary information $index_{s}=\{(x,y,frame),{\rm Video~ID}\}$ is also found according to the established video index database.
\par
\begin{figure}[!t]
	\centering
	\includegraphics[width=0.5\textwidth]{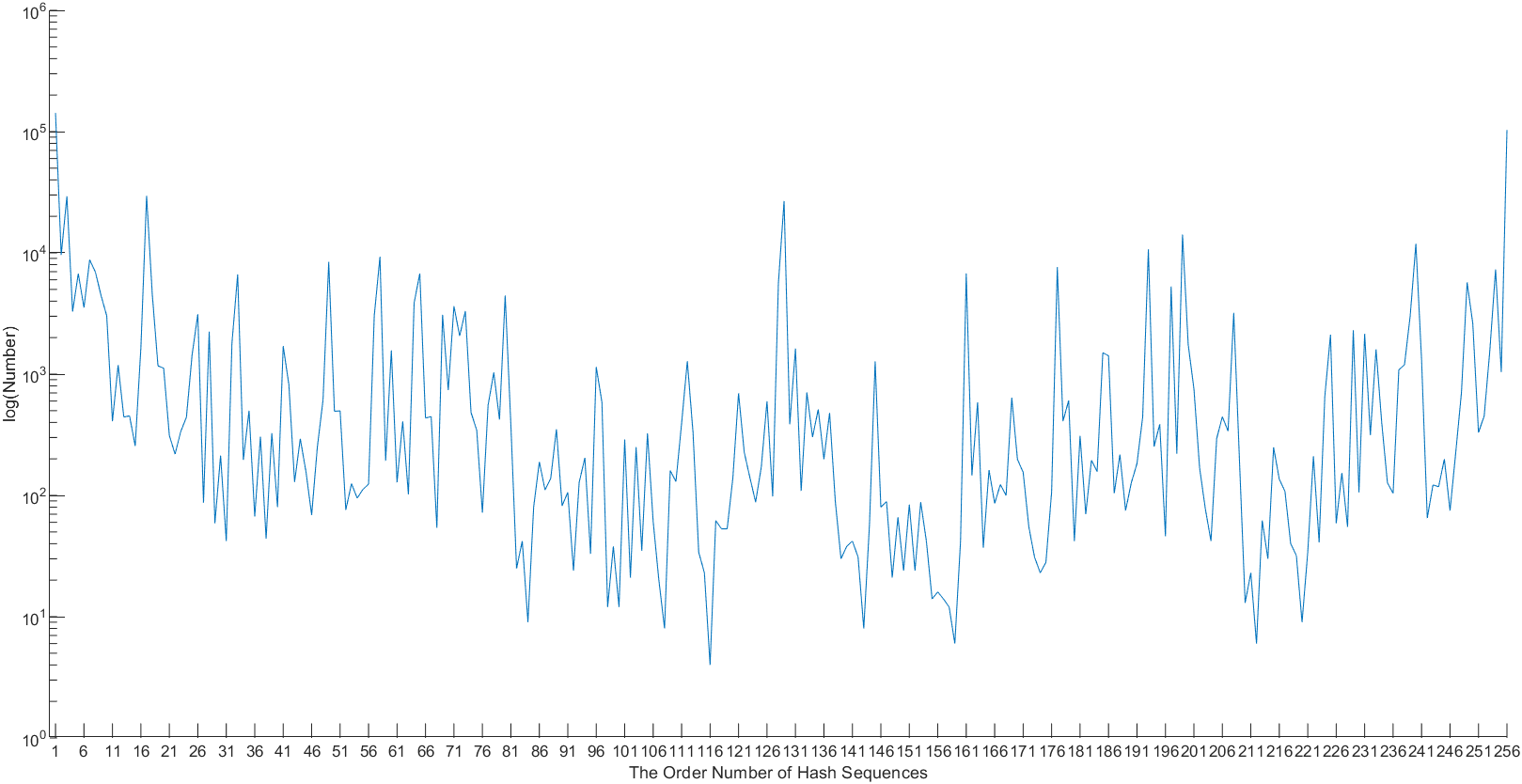}
	\caption{The distribution of the generated hash sequences for the video \emph{PeopleOnStreet}.}
	\label{fig_5}
\end{figure}

 Specifically, the selection rule for videos is as follows. To maximize the use of videos, for the reappeared secret information segments in the same transmission, the same branch will be chosen. Besides, for different secret information segments, it tends to choose the same video, which can be guaranteed by the fact that a video can generate almost all types of hash sequences. Fig. \ref{fig_5} shows the distribution of the generated hash sequences for the video \emph{PeopleOnStreet}, which contains the entire 256 types. Note that the horizontal coordinate which ranges from 1 to 256 represents hash sequences from “00000000” to “11111111” in binary incremental order and the vertical coordinate represents the number of each hash sequence after taking the logarithm. To make it clearer, the following example is taken. Assume that the secret information bits in the \emph{n}-th transmission are “00000000, 11111111, 00000000, ……”. For the first segment “00000000”, the location information (1, 1, 1) and video ID \emph{BQMall} are chosen and for the second one “11111111”, it tends to choose the same video ID \emph{BQMall} but different location information (\emph{x}, \emph{y}, \emph{frame}). Then for the third segment, it is the repeated segment “00000000”, so the same video ID \emph{BQMall} and location information (1, 1, 1) are chosen. Considering the security of transmission, for the reappeared secret information segments in different transmissions, different videos will be chosen. For example, assume that the segment “00000000” reappears in the \emph{n}+1-th transmission, the chosen video \emph{BQMall} in the \emph{n}-th transmission will be avoided as much as possible. At the same time, the corresponding location information and video ID will be sent to the receiver as the corresponding auxiliary information.
\par \emph{Step} 3. Repeat \emph{Step} 2 until all cover videos for the secret information are obtained. Then the location information and video ID which have been discussed in Section \uppercase\expandafter{\romannumeral4}.\emph{B} are recorded as one part of auxiliary information.
\par \emph{Step} 4. The number of padding zeros (If exist.) is recorded as another part of auxiliary information.
\par \emph{Step} 5. The selected videos and the corresponding auxiliary information are sent to the receiver. To make auxiliary information more secure, AES encryption algorithm is adopted.
\subsection{Secret Information Extraction}
\par There may be a variety of attacks on cover videos during the process of transmission, such as the geometric attack, noise attack, frame deletion attack and compression attack. So, the robustness of the scheme depends on whether the secret information can be successfully extracted. For the receiver, the specific steps of extracting secret information are as follows.
\par \emph{Step} 1. The auxiliary information including location information and video ID is obtained by AES decryption algorithm.
\par \emph{Step} 2. The video and corresponding frame $\bm{F}$ are obtained according to the location information and video ID.
\par \emph{Step} 3. The corresponding frame $\bm{F}$ is partitioned into \emph{m}$\times$\emph{n} blocks, which can be described as
\begin{equation}
\label{eqn_16}
\boldsymbol{Y}_{b}=\left\{\boldsymbol{Y}_{(1,1)}, \boldsymbol{Y}_{(1,2)}, \ldots, \boldsymbol{Y}_{(m, n)} \right\} .
\end{equation}
\par \emph{Step} 4. The corresponding block $\bm{Y}_{x,y}$ is obtained according to the coordinate (\emph{x}, \emph{y}). Then Y channel of the block is extracted and partitioned into 16 sub-blocks, which can be obtained by raster scan order
\begin{equation}
\label{eqn_17}
\bm{y}_{i}=\{\bm{y}_{1},\bm{y}_{2},\ldots, \bm{y}_{16}\}, 0<i\le 16 .
\end{equation}
\par \emph{Step} 5. 8$\times$8 DCT is applied for each sub-block $\bm{y}_{i}$ and $\bm{DC}_{i}$ is obtained by adding the DC coefficients of all 8$\times$8 blocks in $\bm{y}_{i}$
\begin{equation}
\label{eqn_18}
\bm{DC}_{i}=\sum_{j=1}^{\left(\frac{M \times N}{32 m \times 32 n}\right)} \bm{C}_{j}(0,0) .
\end{equation}
\par \emph{Step} 6. The maximum of DC coefficients $\bm{DC}_{i}$ is obtained by
\begin{equation}
\label{eqn_19}
\boldsymbol{DC}_{max}=\max _{i}\left\{DC_{1}, DC_{2}, \ldots, DC_{i}, \ldots, DC_{16}\right\} .
\end{equation}
\par \emph{Step} 7. Each bit of hash sequence $\bm{\!H\!S}_{(x,y)}$ for the block $\bm{Y}_{(x,y)}$ is calculated by
\begin{equation}
\label{eqn_20}
\bm{HS}_{(x, y)}^{i}=\left\{\begin{array}{ll}
1,&if \frac{\bm{DC}_{i}}{\bm{DC}_{\max }}>T \vspace{1ex}\\
0,&if \frac{\bm{DC}_{i}}{\bm{DC}_{\max }} \leq T
\end{array}, 0<i<16 .\right.
\end{equation}
\par \emph{Step} 8. The above steps are repeated until all the hash sequences are extracted. By connecting hash sequences and subtracting padding zeros, the secret information $\bm{B'}$  is obtained. Note that the number of cover videos maybe not equal to \emph{S}, which means that a cover video corresponds to at least two secret information segments.
\section{Experimental Results and Analysis}
\subsection{Configuration}
\par Experiments are implemented in the following configuration: Intel(R) Core (TM) i5-10400F CPU @ 2.90GHz, 32GB memory, MATLAB R2019a. Some attacks such as noise and filtering are realized by the library function of MATLAB. In the video compression, the latest high-performance and widely used video coding standard, HEVC, is considered. The process of encoder is implemented by an open source software x265, and the platform for HEVC decoding is HM16.15. In the process of compression, the GOP size is 4 with IPPP coding structure and the value of $QP$ is 22, 27, 32, and 37, respectively. The datasets include DAVIS-2017 dataset and HEVC standard testing sequences, which are described as follows.
\par (1) The DAVIS-2017 dataset. The DAVIS dataset was provided by the DAVIS Challenge \cite{29}. It is a high-quality dataset, which include 90 video sequences in different scenes. The resolution we use is 480p, which is the resolution of official metrics.
\par (2) The HEVC standard test sequences dataset. This dataset includes 18 YUV sequences with different resolutions, which is collected from the standard HEVC test database (http://ftp.kw.bbc.co.uk/hevc/hm-10.0-anchors/bitstreams/). There are five classes videos, Class A to Class E, in total, of which resolutions are 2560$\times$1600, 1920$\times$1080, 832$\times$480, 416$\times$240, and 1280$\times$720, respectively. The number of videos in each class is 2, 5, 4, 4, and 3, respectively.
\par Some pre-designed parameters are discussed as follows. The length of hash sequence \emph{L} is set from 1 to 15. As mentioned before, the threshold \emph{T} is set to meet the ratio of bits 0 and 1 in hash sequence close to 1:1, which is related to specific videos and corresponding resolutions. Through experimental testing, the value of \emph{T} is around 0.85, with a fluctuation of 0.1. Note that the video frame size of DAVIS-2017, Class B, Class D, and Class E is cropped to generate different lengths of hash sequences. From (\ref{eqn_3})-(\ref{eqn_7}), it can be derived that the width and height of the video frame must be an integer multiple of 32. For example, for Class B, the original resolution of the video frame is 1920$\times$1080, so it will be cropped to 1920$\times$1056 when it is used for generating hash sequences. After that, different block sizes for different video classes shown in Table \ref{tab4} and \ref{tab5} can be obtained, in which the block size can be calculated by dividing the cropped resolution by the number of partitioning blocks.
\par The experimental results are compared with the existing coverless steganography algorithms \cite{25,26,27}, where the algorithms of \cite{25,26} are the state-of-the-art coverless image steganography algorithms, and the work of \cite{27} is the only published coverless video steganography. Among them, \cite{25,26} are implemented again in our dataset and the data in \cite{27} is used directly for comparison. In the experimental results, three metrics are mainly considered, which are capacity, robustness, and security, respectively. 
\subsection{Capacity}
\par The capacity of the existing coverless steganography based on mapping rules depends on the length of hash sequences \emph{L}. However, oversized \emph{L} will lead to the inefficiency of searching in index database. So, we set \emph{L} from 1 to 15. In the coverless image steganography, the capacity is judged by the number of images needed when the same information is hidden, which is also called ideal capacity. The specific comparison result is shown in Table \ref{tab1}. As seen from Table \ref{tab1}, the ideal capacity of the proposed algorithm is close to \cite{25,26}. The difference is that \cite{25,26} use an additional image to record the number of padding zeros, while the proposed algorithm transmits this part as auxiliary information, which is consistent with Pan’s \cite{27} algorithm. Obviously, it is not appropriate to transmit an additional image for the coverless video steganography. As for Pan’s \cite{27} algorithm, the ideal capacity range is a subset of the above three algorithms.
\begin{table}[htbp]
	\centering
	\caption{Ideal Capacity Comparison}
	\begin{tabular}{p{5.2em}<{\centering}p{4em}<{\centering}p{3.5em}<{\centering}p{3.5em}<{\centering}p{4em}<{\centering}p{2.5em}<{\centering}}
		\toprule
		\multirow{2}[4]{*}{Algorithm} & \multicolumn{4}{c}{Length of hidden data} & \multirow{2}[4]{*}{\textit{L}} \\
		\cmidrule{2-5}    \multicolumn{1}{c}{} & 1B    & 10B   & 100B  & 1KB   & \multicolumn{1}{c}{} \\
		\midrule
		Zhang’s \cite{25} & 2-9   & 7-81  & 55-801 & 548-8193 & 1-5 \\
		Liu’s \cite{26} & 2-9   & 7-81  & 55-801 & 548-8193 & 1-15 \\
		Pan’s \cite{27} & 1 & 6-10  & 54-100 & 547-1024 & 8-15 \\
		Ours  & 1-8   & 6-80  & 54-800 & 547-8192 & 1-15 \\
		\bottomrule
	\end{tabular}%
	\label{tab1}%
\end{table}%
\par In fact, if the ideal capacity is achieved, it means that the hash sequences generated by each frame are different. For images, it can be implemented because the content between different images changes significantly. However, since the changes in content between adjacent video frames are often slow, the ideal capacity cannot be reached in most cases of videos. Based on this fact, Pan et al. \cite{27} introduced the effective capacity to further evaluate the actual capacity, which is defined as follows.
\par Assume that $\bm{\!H\!S}_{i}=\{hs_{1},hs_{2},\ldots,hs_{8}\}$ is the hash sequence generated by the \emph{i}-th frame. The decimal conversion is done firstly by
\begin{equation}
\label{eqn_21}
D_{\bm{\!H\!S}_{i}}=\sum_{j=1}^{8} h s_{j} \times 2^{8-j} ,
\end{equation}
where $D_{\bm{\!H\!S}_{i}}$ represents the decimal value that the hash sequence of the \emph{i}-th frame is converted into. Then different decimal values of the whole video $\bm{D}=\{D_{\bm{\!H\!S}_{1}},D_{\bm{\!H\!S}_{2}},\ldots,D_{\bm{\!H\!S}_{p}}\}$ is recorded, where \emph{p} represents the number of different decimal values, and the effective capacity is obtained by
\begin{equation}
\label{eqn_22}
C_{E}=\sum_{k=0}^{255} f(k), f(k)=\left\{\begin{array}{l}
1, if~k~in~\boldsymbol{D} \vspace{1ex}\\
0, Otherwise 
\end{array} .\right.
\end{equation}
\begin{table}[htbp]
	\centering
	\caption{The Effective Capacity in DAVIS-2017 Dataset}
	\begin{tabular}{p{12.5em}<{\centering}ccccc}
		\toprule
		Length of hash sequence & \multicolumn{1}{p{2em}}{8bits} & \multicolumn{1}{p{2em}}{9bits} & \multicolumn{1}{p{2em}}{10bits} & \multicolumn{1}{p{2em}}{11bits} & \multicolumn{1}{p{2em}}{12bits} \\
		\midrule
		Theoretical maximum value & 256   & 512   & 1024  & 2048  & 4096 \\
		Pan’s \cite{27} & 242   & 422   & 772   & 1222  & 1752 \\
		Ours ($m=13$, $n=5$) & \textbf{\underline{256}} & \textbf{\underline{512}} & 1020  & 1997  & 3782 \\
		Ours ($m=26$, $n=5$) & \textbf{\underline{256}} & \textbf{\underline{512}} & 1020  & 2002  & 3801 \\
		Ours ($m=13$, $n=15$) & \textbf{\underline{256}} & \textbf{\underline{512}} & \textbf{\underline{1024}} & \textbf{\underline{2048}} & 4078 \\
		Ours ($m=26$, $n=15$) & \textbf{\underline{256}} & \textbf{\underline{512}} & \textbf{\underline{1024}} & \textbf{\underline{2048}} & \textbf{4093} \\
		\bottomrule
	\end{tabular}%
	\label{tab2}%
\end{table}%
\par Thus, the effective capacity represents how many different hash sequences can be generated in the entire video. It is obvious to see that the effective capacity is related to the hash sequence generation method and the video content, and the maximum effective capacity of 8-bit hash sequence is 256. The experimental result on DAVIS-2017 dataset is shown in Table \ref{tab2}. The best results are shown in bold and the desired results in best results are underlined. As seen from Table \ref{tab2} , the effective capacity of the proposed algorithm outperforms that of Pan’s \cite{27} algorithm for all lengths of hash sequences. There is a gap between Pan’s \cite{27} algorithm and the theoretical maximum value. Especially when the length of hash sequence is larger than 10 bits, Pan’s \cite{27} algorithm only reached about 50$\%$ of the theoretical maximum value. In Pan’s \cite{27} algorithm, the semantic segmentation is done for each frame and the hash sequence is generated by comparing the adjacent semantic information proportion, which may generate similar hash sequences for situations where the background occupies large proportion. Whereas for the proposed algorithm, the hash sequence has no significantly relationship with whether the corresponding position is an object or background, so it can generate different hash sequences for the background. In fact, the selection of features and the new hash sequence generation method in the proposed steganography bring richer hash sequences and achieve larger effective capacity together. 
\par Besides, the way of block partitioning also affects the effective capacity. The effective capacity increases as the number of partitioning blocks increases. There are two reasons. First, when the number of partitioning blocks increases, the number of hash sequences that can be generated by a video frame increases. Moreover, more partitioning blocks always mean a smaller block size. When the block is further partitioned into 16 sub-blocks and taken 8$\times$8 DCT, the number of DC coefficients for each sub-block used to generate the average  value of DC coefficients, $\bm{DC}_{i}$, is less, which indicates that the probability of a different relationship between the adjacent $\bm{DC}_{i}$ is larger. Consequently, it makes hash sequences diverse. In most cases, it can reach the maximum desired value when the block size is relatively small as shown in Table \ref{tab2}.

\par For the existing coverless steganography, auxiliary information needs to be transmitted to the receiver along with the carriers, which is some kind of cost. However, the existing steganography algorithms always ignored this important issue. If the length of the auxiliary information is too long compared with the secret one, it will lose the meaning of coverless steganography. So, a new metric, relative effective capacity is proposed for the first time in this paper, which is defined as
\begin{equation}
\label{eqn_23}
C_{RE}=\frac{C_{E}}{L_{a}} ,
\end{equation}
where $C_{E}$ is the effective capacity and $L_{a}$ is the number of binary bits for the auxiliary information. For example, if the range of the video ID is 1 to 90, it will denote $L_{a}=7$ bits for the auxiliary information. The experimental result of the relative effective capacity on DAVIS-2017 dataset is shown in Table \ref{tab3}. We can see that when the length of the hash sequence is 8 bits and 9 bits, Pan’s \cite{27} algorithm is comparable to ours. However, as the length of the hash sequences increases, the gap between Pan’s algorithm and the proposed algorithm gradually increases, especially when the length achieves 12 bits, the relative effective capacity of the proposed algorithm is almost twice that of Pan’s algorithm. Besides, the relative effective capacity of the proposed algorithm fluctuates slightly in different parameters. Overall, under the parameter $m=13$  and $n=5$ , it achieves the best balance between $C_{E}$ and $L_{a}$.

\par In short, with the proposed new hash sequence generation method, the proposed coverless steganography performs better compared with the existing coverless video steganography from ideal capacity which is the universal metric for images and videos, effective capacity and relative effective capacity which are the unique metrics for videos. What’s more, the result achieves the theoretical maximum value in some cases.
\begin{table}[htbp]
	\centering
	\caption{The Relative Effective Capacity in DAVIS-2017 Dataset}
	\begin{tabular}{p{12.5em}<{\centering}ccccc}
		\toprule
		Length of hash sequence & \multicolumn{1}{p{2em}}{8bits} & \multicolumn{1}{p{2em}}{9bits} & \multicolumn{1}{p{2em}}{10bits} & \multicolumn{1}{p{2em}}{11bits} & \multicolumn{1}{p{2em}}{12bits} \\
		\midrule
		Pan’s \cite{27} & 12.1   & 21.1   & 38.6   & 61.1  & 87.6 \\
		Ours ($m=13$, $n=5$) & \textbf{12.2} & \textbf{24.4} & \textbf{48.6}  & \textbf{95.1}  & 180.1 \\
		Ours ($m=26$, $n=5$) & 11.6 & 23.3 & 46.4  & 91.0  & 172.8 \\
		Ours ($m=13$, $n=15$) & 11.6 & 23.3 & 46.5 & 93.1 & \textbf{185.4} \\
		Ours ($m=26$, $n=15$) & 11.1 & 22.3 & 44.5 & 89.0 & 178.0 \\
		\bottomrule
	\end{tabular}%
	\label{tab3}%
\end{table}%
\subsection{Robustness}
\par In coverless steganography algorithms, robustness is an important performance metric, which directly determines whether the secret information can be extracted correctly from the carriers after attacks. In this section, the DAVIS-2017 dataset and HEVC standard test sequences dataset are considered to test the robustness of the proposed algorithm. Note that in order to further verify the performance at different resolutions, the HEVC standard test sequences dataset is utilized to compare with \cite{25,26}. While DAVIS-2017 dataset is consistent with Pan’s \cite{27} experiment. Multiple types of traditional attacks are used to the cover videos, which can be seen from the specific comparison results in Table \ref{tab4}and Table \ref{tab5}. Note that C-H-E means color histogram equalization. Besides, another common but important attack, the frame deletion attack, is considered in our experiment, which will be discussed in Part.2 specifically.
\par Suppose that each hash sequence generated by each block of the video frames is $\bm{\!H\!S}_{i}$, the number of hash sequences generated by the whole dataset is \emph{W}. After attack, the hash sequence is denoted as $\bm{\!H\!S'}_{i}$ and the extraction accuracy is defined as
\begin{equation}
\label{eqn_24}
\begin{aligned}
ACC &=\frac{\sum_{i=1}^{W} f(i)}{W} \!\times\! 100 \%,\\
f(i)&=\left\{\begin{array}{l}
1, if~\boldsymbol{\!H\!S}_{i}=\boldsymbol{\!H\!S}_{i}^{\prime} \vspace{1ex}\\
0, if~\boldsymbol{\!H\!S}_{i} \neq \boldsymbol{\!H\!S}_{i}^{\prime}
\end{array} .\right.
\end{aligned}
\end{equation}
\par It is worth noting that the length of hash sequence \emph{L} is set to 8 in robustness experiments. The specific results of the experiment will be discussed in the following parts referring to the traditional attacks and the frame deletion attack.
\begin{table*}[htbp]
	\centering
	\caption{Extraction Accuracy of Our Algorithm with Different Block Sizes in Class A, Class B, and Class C}
	\begin{tabular}{p{8.04em}lllll|lll|llll}
		\toprule
		\multirow{2}[2]{*}{Attack} & \multicolumn{1}{l}{\multirow{2}[2]{*}{Block size}} & \multicolumn{4}{p{16em}<\centering|}{Class A} & \multicolumn{3}{p{12em}<\centering|}{Class B} & \multicolumn{4}{p{16em}<\centering}{Class C}\\
		\cmidrule{3-13}    \multicolumn{1}{c}{} &       & \multicolumn{1}{l}{32$\times$32} & \multicolumn{1}{l}{64$\times$32} & \multicolumn{1}{l}{64$\times$64} & \multicolumn{1}{l|}{128$\times$64} & \multicolumn{1}{l}{32$\times$32} & \multicolumn{1}{l}{64$\times$96} & \multicolumn{1}{l|}{128$\times$96} & \multicolumn{1}{l}{32$\times$32} & \multicolumn{1}{l}{64$\times$32} & \multicolumn{1}{l}{32$\times$96} & \multicolumn{1}{l}{64$\times$96}\\
		\midrule
		\multirow{4}[1]{*}{HEVC compression} & \multicolumn{1}{l}{$QP$(22)} & 96.5\% & 97.6\% & 98.2\% & \textbf{98.7\%} & 93.8\% & 97.9\% & \textbf{98.4\%} & 93.8\% & 96.9\% & 97.5\% & \textbf{98.0\%}\\
		\multicolumn{1}{l}{} & \multicolumn{1}{l}{$QP$(27)} & 94.4\% & 95.9\% & 97.0\% & \textbf{97.8\%} & 90.4\% & 96.6\% & \textbf{97.5\%} & 90.0\% & 94.8\% & 95.9\% & \textbf{96.7\%}\\
		\multicolumn{1}{l}{} & \multicolumn{1}{l}{$QP$(32)} & 91.4\% & 93.5\% & 95.1\% & \textbf{96.3\%} & 85.8\% & 94.5\% & \textbf{95.9\%} & 84.3\% & 91.5\% & 93.5\% & \textbf{94.8\%}\\
		\multicolumn{1}{l}{} & \multicolumn{1}{l}{$QP$(37)} & 87.2\% & 89.8\% & 92.2\% & \textbf{93.9\%} & 79.6\% & 91.2\% & \textbf{93.3\%} & 76.8\% & 86.7\% & 89.5\% & \textbf{91.4\%} \\
		\multirow{3}[0]{*}{Gauss noise} & \multicolumn{1}{l}{$\sigma$(0.001)} & 91.5\% & 93.8\% & 95.7\% & \textbf{96.5\%} & 87.3\% & 96.1\% & \textbf{97.2\%} & 83.8\% & 91.9\% & 93.9\% & \textbf{95.3\%}\\
		\multicolumn{1}{l}{} & \multicolumn{1}{l}{$\sigma$(0.005)} & 83.6\% & 87.6\% & 91.3\% & \textbf{92.7\%} & 73.1\% & 91.1\% & \textbf{94.1\%} & 68.1\% & 83.3\% & 87.4\% & \textbf{90.5\%}\\
		\multicolumn{1}{l}{} & \multicolumn{1}{l}{$\sigma$(0.1)} & 45.9\% & 58.7\% & \textbf{66.1\%} & 64.3\% & 29.9\% & 69.3\% & \textbf{74.3\%} & 27.4\% & 43.1\% & 54.6\% & \textbf{65.7\%}\\
		\multirow{3}[0]{*}{Salt\&pepper noise} & \multicolumn{1}{l}{$\sigma$(0.001)} & 97.9\% & 98.0\% & 98.3\% & \textbf{98.4\%} & 96.5\% & 98.3\% & \textbf{98.5\%} & 96.2\% & 97.6\% & 97.9\% & \textbf{98.0\%}\\
		\multicolumn{1}{l}{} & \multicolumn{1}{l}{$\sigma$(0.005)} & 92.0\% & 93.2\% & 95.0\% & \textbf{95.7\%} & 86.9\% & 95.4\% & \textbf{96.6\%} & 85.4\% & 91.6\% & 93.3\% & \textbf{94.7\%}\\
		\multicolumn{1}{l}{} & \multicolumn{1}{l}{$\sigma$(0.1)} & 65.0\% & 72.7\% & \textbf{77.1\%} & 76.7\% & 45.2\% & 79.7\% & \textbf{83.1\%} & 40.7\% & 64.8\% & 73.0\% & \textbf{77.7\%}\\
		\multirow{3}[0]{*}{Speckle noise} & \multicolumn{1}{l}{$\sigma$(0.01)} & 91.0\% & 93.1\% & 95.0\% & \textbf{95.7\%} & 85.4\% & 95.1\% & \textbf{96.3\%} & 77.4\% & 88.4\% & 91.2\% & \textbf{93.1\%}\\
		\multicolumn{1}{l}{} & \multicolumn{1}{l}{$\sigma$(0.05)} & 81.6\% & 85.3\% & 88.2\% & \textbf{88.9\%} & 65.7\% & 89.2\% & \textbf{91.4\%} & 55.9\% & 76.0\% & 81.9\% & \textbf{85.5\%}\\
		\multicolumn{1}{l}{} & \multicolumn{1}{l}{$\sigma$(0.1)} & 74.6\% & 79.5\% & 82.8\% & \textbf{83.3\%} & 49.9\% & 85.0\% & \textbf{87.5\%} & 43.4\% & 67.1\% & 75.0\% & \textbf{80.1\%}\\
		\multirow{3}[0]{*}{Median filtering} & \multicolumn{1}{l}{(3$\times$3)} & 96.1\% & 97.0\% & 97.8\% & \textbf{98.1\%} & 93.2\% & 97.2\% & \textbf{97.5\%} & 87.9\% & 92.5\% & 93.7\% & \textbf{94.5\%}\\
		\multicolumn{1}{l}{} & \multicolumn{1}{l}{(5$\times$5)} & 90.1\% & 92.0\% & 94.3\% & \textbf{94.9\%} & 86.8\% & 93.9\% & \textbf{94.6\%} & 78.5\% & 86.2\% & 88.4\% & \textbf{89.4\%}\\
		\multicolumn{1}{l}{} & \multicolumn{1}{l}{(7$\times$7)} & 84.1\% & 86.9\% & 90.5\% & \textbf{91.6\%} & 81.2\% & 91.0\% & \textbf{92.0\%} & 70.8\% & 81.3\% &83.8\% & \textbf{85.7\%}\\
		\multirow{3}[0]{*}{Mean filtering} & \multicolumn{1}{l}{(3$\times$3)} & 95.2\% & 96.4\% & 97.6\% & \textbf{98.1\%} & 93.0\% & 97.5\% & \textbf{97.7\%} & 86.5\% & 92.0\% & 93.7\% & \textbf{94.5\%}\\
		\multicolumn{1}{l}{} & \multicolumn{1}{l}{(5$\times$5)} & 89.1\% & 91.4\% & 94.5\% & \textbf{95.5\%} & 86.2\% & 94.8\% & \textbf{95.5\%} & 76.1\% & 85.5\% & 88.1\% & \textbf{89.9\%}\\
		\multicolumn{1}{l}{} & \multicolumn{1}{l}{(7$\times$7)} & 82.6\% & 86.1\% & 91.0\% & \textbf{92.4\%} & 79.3\% & 92.1\% & \textbf{93.4\%} & 66.5\% & 78.9\% & 82.2\% & \textbf{85.6\%}\\
		\multirow{3}[0]{*}{Gauss filtering} & \multicolumn{1}{l}{(3$\times$3)} & 76.5\% & 81.4\% & 87.8\% & \textbf{89.9\%} & 76.6\% & 92.1\% & \textbf{94.7\%} & 64.3\% & 78.2\% & 82.2\% & \textbf{87.7\%}\\
		\multicolumn{1}{l}{} & \multicolumn{1}{l}{(5$\times$5)} & 61.3\% & 67.5\% & 76.4\% & \textbf{80.4\%} & 63.8\% & 85.4\% & \textbf{89.7\%} & 48.0\% & 64.7\% & 70.3\% & \textbf{77.7\%}\\
		\multicolumn{1}{l}{} & \multicolumn{1}{l}{(7$\times$7)} & 51.2\% & 56.2\% & 64.9\% & \textbf{70.3\%} & 55.9\% & 78.9\% & \textbf{84.5\%} & 38.9\% & 54.8\% & 61.2\% & \textbf{68.4\%}\\
		\multirow{2}[0]{*}{Centered cropping} & 20\%  & \textbf{83.6\%} & 83.3\% & 82.2\% & 81.3\% & \textbf{84.1\%} & 82.8\% & 80.4\% & \textbf{80.0\%} & 76.4\% & 77.7\% & 75.3\%\\
		\multicolumn{1}{l}{} & 50\%  & \textbf{59.3\%} & 58.3\% & 56.2\% & 54.7\% & \textbf{60.1\%} & 56.1\% & 54.7\% & \textbf{53.4\%} & 46.8\% & 47.6\% & 43.3\% \\
		\multirow{2}[0]{*}{Edge cropping} & 10\%  & \textbf{91.0\%} & 90.7\% & 88.8\% & 86.1\% & \textbf{89.8\%} & 86.1\% & 82.4\% & \textbf{86.3\%} & 82.3\% & 75.8\% & 72.2\% \\
		\multicolumn{1}{l}{} & 20\%  & \textbf{83.3\%} & 83.1\% & 81.9\% & 81.5\% & \textbf{82.5\%} & 76.4\% & 76.1\% & 59.3\% & \textbf{62.7\%} & 62.7\% & 62.4\%\\
		\multirow{3}[0]{*}{Rotation} & \multicolumn{1}{l}{10°} & \textbf{14.5\%} & 11.2\% & 10.4\% & 7.2\% & 22.4\% & \textbf{23.2\%} & 20.3\% & 11.4\% & 16.2\% & \textbf{22.5\%} & 19.8\% \\
		\multicolumn{1}{l}{} & \multicolumn{1}{l}{30°} & \textbf{12.8\%} & 9.9\% & 9.0\% & 5.6\% & \textbf{16.8\%} & 16.0\% & 12.5\% & 9.3\% & 10.9\% & \textbf{15.7\%} & 12.0\%\\
		\multicolumn{1}{l}{} & \multicolumn{1}{l}{50°} & \textbf{11.7\%} & 8.8\% & 9.2\% & 1.6\% & \textbf{15.1\%} & 14.8\% & 11.8\% & 9.4\% & 9.4\% & \textbf{13.8\%} & 10.7\% \\
		\multirow{3}[0]{*}{Translation} & \multicolumn{1}{l}{(80,50)} & \textbf{15.9\%} & 12.3\% & 11.8\% & 8.5\% & \textbf{21.9\%} & 20.0\% & 16.4\% & 8.2\% & 9.6\% & \textbf{14.5\%} & 12.4\% \\
		\multicolumn{1}{l}{} & (160,100) & \textbf{14.0\%} & 10.4\% & 10.2\% & 6.9\% & \textbf{18.4\%} & 16.7\% & 13.5\% & 9.1\% & 8.1\% & \textbf{11.0\%} & 8.7\% \\
		\multicolumn{1}{l}{} & (320,200) & \textbf{12.6\%} & 9.6\% & 9.5\% & 5.9\% & \textbf{16.2\%} & 14.7\% & 12.3\% & \textbf{11.7\%} & 8.4\% & 11.7\% & 10.5\% \\
		\multirow{5}[0]{*}{Scaling} & 0.3   & 92.9\% & 94.8\% & 96.7\% & \textbf{97.6\%} & 89.3\% & 96.2\% & \textbf{97.0\%} & 76.9\% & 86.1\% & 86.5\% & \textbf{89.7\%} \\
		\multicolumn{1}{l}{} & 0.5   & 97.9\% & 98.6\% & 99.2\% & \textbf{99.4\%} & 96.7\% & 99.2\% & \textbf{99.5\%} & 93.6\% & 96.9\% & 97.8\% & \textbf{98.6\%}\\
		\multicolumn{1}{l}{} & 0.75  & 99.4\% & 99.6\% & 99.8\% & \textbf{99.8\%} & 98.8\% & 99.7\% & \textbf{99.8\%} & 97.4\% & 98.8\% & 99.2\% & \textbf{99.5\%}\\
		\multicolumn{1}{l}{} & 1.5   & 99.7\% & 99.8\% & 99.9\% & \textbf{99.9\%} & 99.2\% & 99.8\% & \textbf{99.9\%} & 98.2\% & 99.2\% & 99.4\% & \textbf{99.6\%}\\
		\multicolumn{1}{l}{} & 3.0     & 99.6\% & 99.8\% & 99.9\% & \textbf{99.9\%} & 99.1\% & 99.8\% & \textbf{99.9\%} & 98.0\% & 99.1\% & 99.4\% & \textbf{99.6\%}\\
		Gamma Correction & 0.8 & \textbf{75.4\%} & 73.3\% & 68.9\% & 64.6\% & \textbf{78.9\%} & 76.7\% & 75.3\% & 68.1\% & 68.9\% & \textbf{72.2\%} & 70.0\% \\
		C-H-E & \multicolumn{1}{l}{None} & \textbf{71.0\%} & 69.8\% & 66.5\% & 63.7\% & 60.9\% & \textbf{63.6\%} & 63.0\% &46.4\% & 47.4\% & \textbf{47.6\%} & 47.0\% \\
		\bottomrule
	\end{tabular}%
	\label{tab4}%
\end{table*}%
\begin{table*}[htbp]
	\centering
	\caption{Extraction Accuracy of Our Algorithm with Different Block Sizes in Class D,Class E and DAVIS-2017}
	\begin{tabular}{p{8.04em}ll|lll|llll}
		\toprule
		\multirow{2}[2]{*}{Attack} & \multicolumn{1}{r}{\multirow{2}[2]{*}{Block size}} & \multicolumn{1}{p{4.04em}|}{Class D} & \multicolumn{3}{c|}{Class E} & \multicolumn{4}{c}{DAVIS-2017} \\
		\cmidrule{3-10}    \multicolumn{1}{r}{} &       & \multicolumn{1}{p{4.04em}|}{32$\times$32} & \multicolumn{1}{p{4.04em}}{32$\times$32} & \multicolumn{1}{p{4.04em}}{64$\times$64} & \multicolumn{1}{p{4.04em}|}{128$\times$64} & \multicolumn{1}{p{4.04em}}{32$\times$32} & \multicolumn{1}{p{4.04em}}{64$\times$32} & \multicolumn{1}{p{4.04em}}{32$\times$96} & \multicolumn{1}{p{4.04em}}{64$\times$96} \\
		\midrule
		\multirow{4}[0]{*}{HEVC compression} & \multicolumn{1}{p{4.04em}}{$QP$(22)} & \textbf{95.7\%} & 90.5\% & 97.0\% & \textbf{98.4\%} & 94.0\% & 96.7\% & 97.4\% & \textbf{98.1\%} \\
		\multicolumn{1}{r}{} & \multicolumn{1}{p{4.04em}}{$QP$(27)} & \textbf{93.2\%} & 86.7\% & 95.4\% & \textbf{97.5\%} & 90.4\% & 94.5\% & 95.7\% & \textbf{96.8\%} \\
		\multicolumn{1}{r}{} & \multicolumn{1}{p{4.04em}}{$QP$(32)} & \textbf{89.0\%} & 82.1\% & 93.0\% & \textbf{96.1\%} & 85.4\% & 91.2\% & 93.2\% & \textbf{94.8\%} \\
		\multicolumn{1}{r}{} & \multicolumn{1}{p{4.04em}}{$QP$(37)} & \textbf{83.4\%} & 76.5\% & 89.0\% & \textbf{93.6\%} & 78.6\% & 86.4\% & 89.3\% & \textbf{91.5\%} \\
		\multirow{3}[0]{*}{Gauss noise} & \multicolumn{1}{p{4.04em}}{$\sigma$(0.001)} & \textbf{91.6\%} & 76.0\% & 93.6\% & \textbf{96.4\%} & 84.6\% & 91.5\% & 93.4\% & \textbf{95.2\%} \\
		\multicolumn{1}{r}{} & \multicolumn{1}{p{4.04em}}{$\sigma$(0.005)} & \textbf{82.3\%} & 51.6\% & 86.5\% & \textbf{92.2\%} & 70.8\% & 82.8\% & 86.6\% & \textbf{90.1\%} \\
		\multicolumn{1}{r}{} & \multicolumn{1}{p{4.04em}}{$\sigma$(0.1)} & \textbf{42.4\%} & 23.7\% & 50.3\% & \textbf{74.0\%} & 27.1\% & 46.9\% & 57.6\% & \textbf{66.0\%} \\
		\multirow{3}[0]{*}{Salt\&pepper noise} & \multicolumn{1}{p{4.04em}}{$\sigma$(0.001)} & \textbf{98.1\%} & 91.9\% & 97.3\% & \textbf{98.4\%} & 96.0\% & 97.4\% & 97.6\% & \textbf{97.9\%} \\
		\multicolumn{1}{r}{} & \multicolumn{1}{p{4.04em}}{$\sigma$(0.005)} & \textbf{92.4\%} & 72.6\% & 92.3\% & \textbf{95.6\%} & 85.5\% & 90.9\% & 92.6\% & \textbf{94.4\%} \\
		\multicolumn{1}{r}{} & \multicolumn{1}{p{4.04em}}{$\sigma$(0.1)} & \textbf{57.9\%} & 29.4\% & 68.9\% & \textbf{82.3\%} & 42.9\% & 64.9\% & 71.8\% & \textbf{76.8\%} \\
		\multirow{3}[0]{*}{Speckle noise} & \multicolumn{1}{p{4.04em}}{$\sigma$(0.01)} & \textbf{87.4\%} & 71.5\% & 93.1\% & \textbf{96.0\%} & 80.3\% & 89.0\% & 91.1\% & \textbf{93.3\%} \\
		\multicolumn{1}{r}{} & \multicolumn{1}{p{4.04em}}{$\sigma$(0.05)} & \textbf{71.3\%} & 39.2\% & 84.4\% & \textbf{91.1\%} & 59.0\% & 77.8\% & 82.1\% & \textbf{85.8\%} \\
		\multicolumn{1}{r}{} & \multicolumn{1}{p{4.04em}}{$\sigma$(0.1)} & \textbf{59.1\%} & 31.9\% & 77.5\% & \textbf{87.2\%} & 44.4\% & 70.2\% & 76.0\% & \textbf{80.3\%} \\
		\multirow{3}[0]{*}{Median filtering} & \multicolumn{1}{p{4.04em}}{(3$\times$3)} & \textbf{91.3\%} & 94.1\% & \textbf{97.8\%} & 97.5\% & 92.0\% & 94.7\% & 95.9\% & \textbf{96.8\%} \\
		\multicolumn{1}{r}{} & \multicolumn{1}{p{4.04em}}{(5$\times$5)} & \textbf{83.7\%} & 89.4\% & 94.7\% & \textbf{95.3\%} & 83.8\% & 88.5\% & 91.2\% & \textbf{92.8\%} \\
		\multicolumn{1}{r}{} & \multicolumn{1}{p{4.04em}}{(7$\times$7)} & \textbf{78.0\%} & 86.1\% & 91.7\% & \textbf{93.7\%} & 77.0\% & 83.0\% & 86.9\% & \textbf{89.2\%} \\
		\multirow{3}[0]{*}{Mean filtering} & \multicolumn{1}{p{4.04em}}{(3$\times$3)} & \textbf{88.0\%} & 89.6\% & 94.2\% & \textbf{96.0\%} & 89.7\% & 93.0\% & 94.2\% & \textbf{95.0\%} \\
		\multicolumn{1}{r}{} & \multicolumn{1}{p{4.04em}}{(5$\times$5)} & \textbf{78.4\%} & 83.2\% & 91.2\% & \textbf{92.4\%} & 80.1\% & 86.3\% & 89.1\% & \textbf{90.9\%} \\
		\multicolumn{1}{r}{} & \multicolumn{1}{p{4.04em}}{(7$\times$7)} & \textbf{70.2\%} & 78.9\% & 87.4\% & \textbf{90.5\%} & 70.9\% & 79.5\% & 83.8\% & \textbf{86.9\%} \\
		\multirow{3}[0]{*}{Gauss filtering} & \multicolumn{1}{p{4.04em}}{(3$\times$3)} & \textbf{76.4\%} & 80.7\% & 92.5\% & \textbf{95.1\%} & 70.1\% & 78.3\% & 84.7\% & \textbf{89.7\%} \\
		\multicolumn{1}{r}{} & \multicolumn{1}{p{4.04em}}{(5$\times$5)} & \textbf{61.6\%} & 69.8\% & 86.0\% & \textbf{90.3\%} & 55.4\% & 65.0\% & 74.0\% & \textbf{80.7\%} \\
		\multicolumn{1}{r}{} & \multicolumn{1}{p{4.04em}}{(7$\times$7)} & \textbf{51.4\%} & 61.9\% & 79.0\% & \textbf{85.1\%} & 46.9\% & 56.0\% & 66.1\% & \textbf{72.2\%} \\
		\multirow{2}[0]{*}{Centered cropping} & 20\%  & \textbf{77.7\%} & \textbf{85.3\%} & 83.7\% & 81.6\% & \textbf{80.9\%} & 76.5\% & 78.2\% & 75.3\% \\
		\multicolumn{1}{r}{} & 50\%  & \textbf{57.8\%} & \textbf{60.4\%} & 57.0\% & 50.6\% & \textbf{54.6\%} & 47.0\% & 47.4\% & 42.3\% \\
		\multirow{2}[0]{*}{Edge cropping} & 10\%  & \textbf{79.6\%} & \textbf{88.5\%} & 82.0\% & 78.5\% & \textbf{87.2\%} & 81.8\% & 79.4\% & 75.0\% \\
		\multicolumn{1}{r}{} & 20\%  & \textbf{75.4\%} & \textbf{79.7\%} & 77.6\% & 71.4\% & 60.5\% & 61.3\% & \textbf{65.5\%} & 65.2\% \\
		\multirow{3}[0]{*}{Rotation} & \multicolumn{1}{p{4.04em}}{10°} & \textbf{23.0\%} & \textbf{27.1\%} & 24.4\% & 22.4\% & 18.2\% & 19.8\% & \textbf{24.5\%} & 21.5\% \\
		\multicolumn{1}{r}{} & \multicolumn{1}{p{4.04em}}{30°} & \textbf{16.2\%} & \textbf{17.8\%} & 14.5\% & 10.4\% & 12.7\% & 12.1\% & \textbf{15.4\%} & 12.4\% \\
		\multicolumn{1}{r}{} & \multicolumn{1}{p{4.04em}}{50°} & \textbf{15.0\%} & \textbf{14.5\%} & 12.0\% & 8.8\% & 11.5\% & 10.0\% & \textbf{13.6\%} & 10.8\% \\
		\multirow{3}[0]{*}{Translation} & \multicolumn{1}{p{4.04em}}{(80,50)} & \textbf{15.5\%} & \textbf{22.0\%} & 19.5\% & 15.6\% & 15.0\% & 14.3\% & \textbf{18.1\%} & 15.4\% \\
		\multicolumn{1}{r}{} & \multicolumn{1}{p{4.04em}}{(160,100)} & \textbf{18.6\%} & \textbf{14.6\%} & 13.5\% & 9.9\% & 13.3\% & 11.0\% & \textbf{14.8\%} & 12.1\% \\
		\multicolumn{1}{r}{} & (320,200) & \textbf{22.8\%} & \textbf{11.6\%} & 9.1\% & 6.0\% & 14.3\% & 10.5\% & \textbf{15.8\%} & 14.2\% \\
		\multirow{5}[0]{*}{Scaling} & 0.3   & \textbf{78.1\%} & 83.6\% & 90.7\% & \textbf{93.7\%} & 82.3\% & 88.0\% & 89.0\% & \textbf{91.9\%} \\
		\multicolumn{1}{r}{} & 0.5   & \textbf{96.1\%} & 96.7\% & 99.2\% & \textbf{99.6\%} & 95.9\% & 97.6\% & 98.5\% & \textbf{99.1\%} \\
		\multicolumn{1}{r}{} & 0.75  & \textbf{98.4\%} & 98.7\% & 99.7\% & \textbf{99.8\%} & 98.5\% & 99.2\% & 99.5\% & \textbf{99.7\%} \\
		\multicolumn{1}{r}{} & 1.5   & \textbf{98.9\%} & 99.2\% & 99.8\% & \textbf{99.9\%} & 99.0\% & 99.5\% & 99.7\% & \textbf{99.8\%} \\
		\multicolumn{1}{r}{} & 3     & \textbf{98.8\%} & 99.1\% & 99.8\% & \textbf{99.9\%} & 98.9\% & 99.4\% & 99.6\% & \textbf{99.8\%} \\
		Gamma Correction & 0.8 & \textbf{81.0\%} & \textbf{83.7\%} & 80.6\% & 79.2\% & 71.6\% & 69.7\% & \textbf{71.9\%} & 69.8\% \\
		C-H-E & \multicolumn{1}{p{4.04em}}{None}  & \textbf{59.6\%} & 60.2\% & 60.1\% & \textbf{61.9\%} & 50.8\% & 47.9\% & \textbf{52.3\%} & 50.3\% \\
		\bottomrule

	\end{tabular}%
	\label{tab5}%
\end{table*}%

\subsubsection{The Traditional Attacks}
\par
\begin{figure*}[!t]
	\centering
	\includegraphics[width=0.9\textwidth]{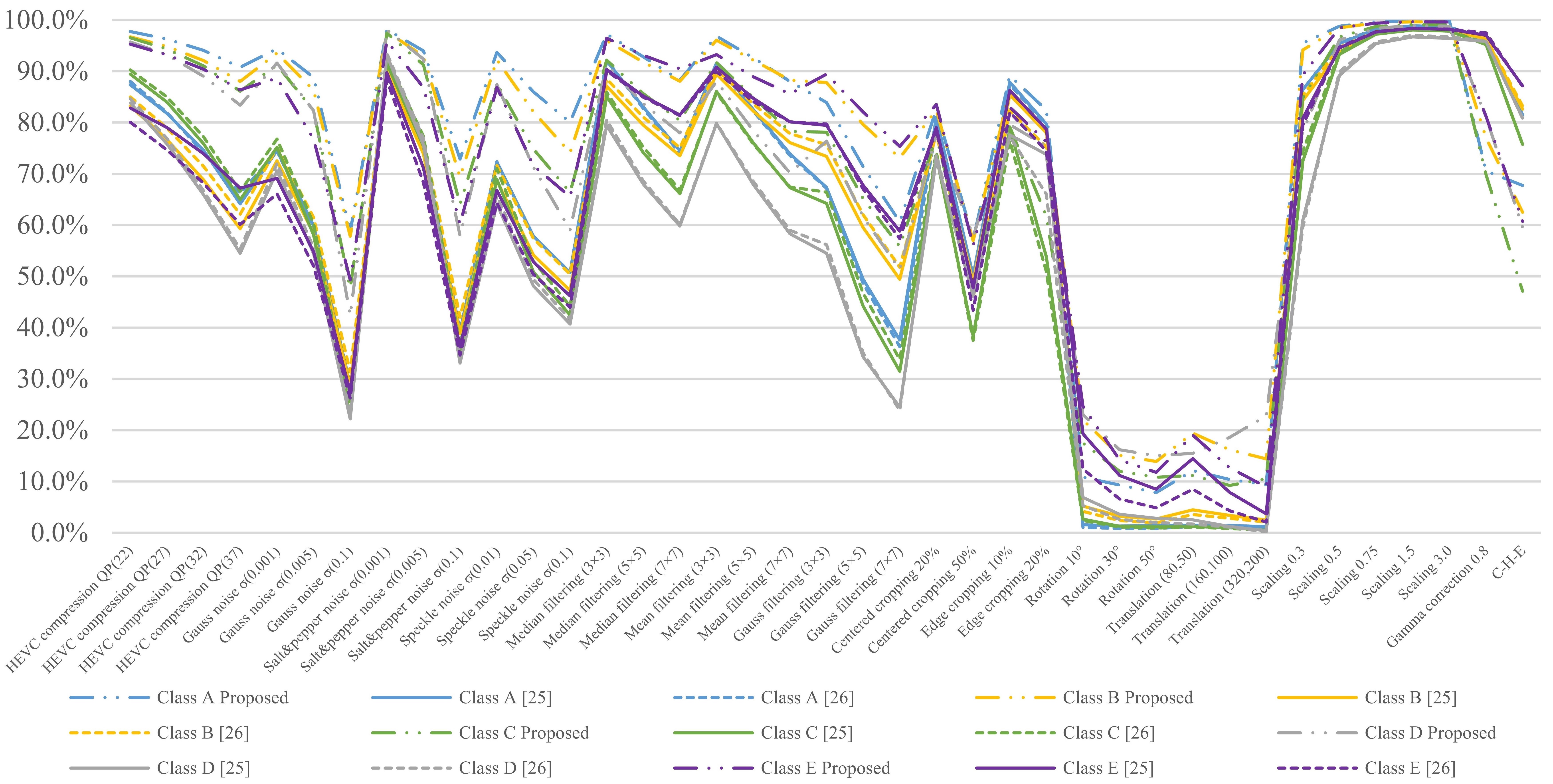}
	\caption{Extraction accuracy of the proposed algorithm with Zhang's \cite{25} and Liu's \cite{26} in the HEVC standard test sequences dataset.}
	\label{fig_6}
\end{figure*}

 In this part, most traditional attacks in \cite{25,26,27} are considered except that JPEG compression is replaced by HEVC compression. Obviously, it is more practical to use video compression rather than image compression for the coverless video steganography. There are two main reasons that we also use video compression for the coverless image steganography. One is that the encoding of I frames in HEVC compression is similar to images in JPEG compression. The other is to test the robustness of hash sequence generation ways in image steganography are strong or not when they are used to resist video compression. The robustness performance of the proposed algorithm on the HEVC standard test sequences dataset and DAVIS-2017 is listed in Table \ref{tab4}and Table \ref{tab5}. Note that only one block partitioning method is considered for Class D due to its low resolution. It can be found that most results are above 80\%, except for the following two cases. One is rotation and translation, which leads to the change of block position and makes hash sequences misaligned. The other is that when the attacks are very severe, for example, when $\sigma = 0.1$ for Gaussian noise, the result is about 50\%, which has serious influence on the frequency domain and the similar failure results are achieved for the existing coverless image steganography \cite{25,26} in these two cases as shown in Fig. \ref{fig_6}. 
\par From the view of different block sizes, it affects the result significantly. On the one hand, the large block size brings better robustness performance against video frame processing attacks and on the other hand, the small block size performs better against content loss attacks. For the video frame processing attacks, such as noise, filtering and compression, there is less loss on video contents. A large block means that more DC coefficients are used to generate a hash sequence, which guarantees the strong robustness. While for the content loss attacks, such as translation and cropping, a small block means that the proportion of the block loss is low, which brings strong robustness. From the view of different resolutions, it can be found that the robustness of high-resolution videos under video frame processing attacks outperforms that of low-resolution ones. For the high-resolution videos, the number of partitioning blocks is larger compared with the low-resolution ones under the same block size. If a hash sequence generated by a block is wrong under some attack, the impact on the extraction accuracy is smaller. From the view of different datasets, it can be seen that the result of the proposed algorithm  in DAVIS-2017 is similar to that of the HEVC standard test sequences dataset, which indicates that the proposed algorithm can be adapted to different datasets.
\par The comparison result with Zhang’s \cite{25} and Liu’s \cite{26} algorithms is shown in Fig. \ref{fig_6}. Note that the result is the average of the different block sizes for each Class. It can be found that the extraction accuracy of the proposed algorithm has increased by more than 10 percentage points under most attacks. Especially for video compression, when the value of $QP$ is small, \cite{25,26} can resist video compression to certain extent. However, when $QP$ is large, such as $QP=37$ , the results of \cite{25,26} can only be kept around 70\% while most results of the proposed algorithm are still achieved around 90\%, which indicates that it cannot transplant image hash sequence generation ways into coverless video steganography directly. The main reason is that the distribution of DC maximum coefficients adopted in this paper is more stable than that of adjacent DC coefficients used in \cite{25,26}, which has been discussed in Section \uppercase\expandafter{\romannumeral3}. For gamma correction and color histogram equalization, Zhang’s \cite{25} and Liu’s \cite{26} perform better. However, as a matter of fact, these are attacks of image processing, which are uncommon in videos. 
\par Meanwhile, comparison is done between the proposed algorithm and Pan’s \cite{27} algorithm. For fair comparison, the types of attacks and parameter settings are consistent with Pan’s work in which only JPEG compression instead of video compression is tested and the result is listed in Table \ref{tab6}. It can be found that the robustness of proposed algorithm is better than Pan’s algorithm under various parameters. Most results improve more than 20 percentage points, especially the increasement of 50 percentage points under Gauss noise attack. The specific reason that the robustness of Pan’s \cite{27} algorithm is not strong may be the insufficient depth and breadth of the network segmentation, which has been described in \cite{27}.

\par In short, it is worth emphasizing that it is necessary to consider the robustness of resisting video compression in coverless video steganography. However, the existing algorithms have not considered about it, and it is the first time to consider video compression attack in this paper. Besides, a new hash sequence generation method based on the maximum DC coefficients is proposed, which brings stronger robustness in different video datasets, parameter settings, and attacks compared with the existing algorithms.
\begin{table}[htbp]
	\centering
	\caption{Extraction Accuracy of The Proposed Algorithm with Pan's \cite{27} in DAVIS-2017 Dataset}
	\begin{tabular}{p{12em}cc}
		\toprule
		Attack & \multicolumn{1}{l}{Pan’s \cite{27}} & \multicolumn{1}{l}{Ours} \\
		\midrule
		Salt\&pepper noise($\sigma$=0.001) & 83.3\% & \textbf{97.9\%} \\
		Salt\&pepper noise($\sigma$=0.005) & 61.7\% & \textbf{94.4\%} \\
		Gauss noise($\sigma$=0.001) & 30.6\% & \textbf{95.2\%} \\
		Gauss noise($\sigma$=0.005) & 29.8\% & \textbf{90.1\%} \\
		Speckle noise($\sigma$=0.01) & 49.6\% & \textbf{93.3\%} \\
		Speckle noise($\sigma$=0.05) & 30.0\% & \textbf{85.8\%} \\
		JPEG compression($Q$(70)) & 77.2\% & \textbf{98.3\%} \\
		JPEG compression($Q$(90)) & 88.3\% & \textbf{99.4\%} \\
		\bottomrule
	\end{tabular}%
	\label{tab6}%
\end{table}%

\subsubsection{The frame deletion attack}
\par In this part, the frame deletion attack is discussed. The frame deletion attack is a common attack in the time domain of videos, and it will cause disorder of video frame numbers resulting in wrong extraction of secret bits. Therefore, it is also important to make coverless video steganography robust against frame deletion as video compression.
\par
\begin{figure}[!t]
	\centering
	\includegraphics[width=0.5\textwidth]{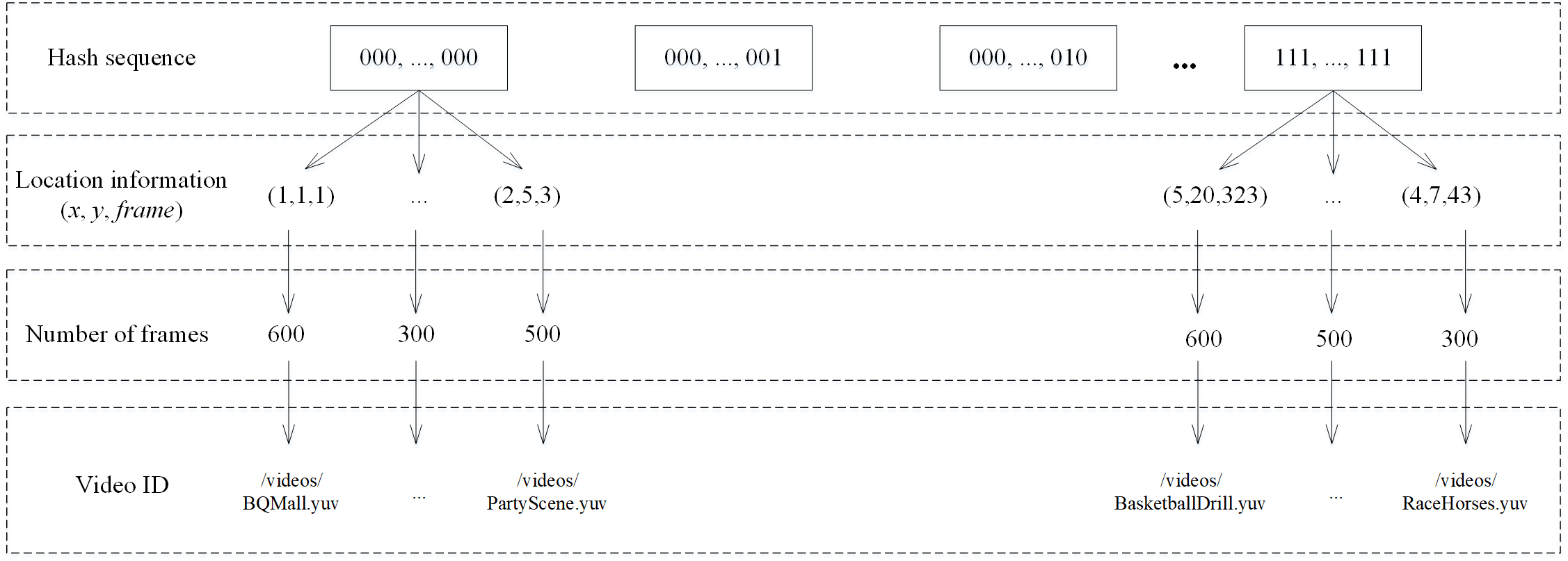}
	\caption{The new structure of video index database.}
	\label{fig_7}
\end{figure}

 Firstly, a video-frame deletion detection based on consistency of quotients of MSSIM (Mean of Structural Similarity) \cite{30} is applied to locate the deleted frames in this paper. The quotients of MSSIM for adjacent frames is calculated to extract features, and the Chebyshev’s inequality is used twice to detect abnormal points, then the deleted frames can be located. Secondly, the structure of the video index database discussed in Section \uppercase\expandafter{\romannumeral4} needs to add a new part ‘Number of frames’ to guarantee the correct orders of the video frames after the frame deletion, and the new one can be seen in Fig. \ref{fig_7}. Although the cost of auxiliary information will increase a little, it can significantly improve the ability to resist frame deletion attack. The experiment is done on Class B and Class E of the HEVC standard test sequences dataset, respectively. Since the frame deletion attack deletes the entire video frame, different block partitioning methods will bring out similar results of extraction accuracy as for different block partitioning methods in other attacks. So, only one block partitioning method is considered in the experiment. In Class B, $m = 15$  and $n = 11$  while in Class E, $m = 10$  and $n = 11$ . The result is shown in Table \ref{tab7}. It can be found that the extraction accuracy is around 90\%. As a matter of fact, if the frame deletion location is correctly detected, the hash sequences can be extracted correctly except for the deleted frames. Thus, the proposed algorithm can resist to small number of frames deletion very well. For the severe frame deletion which means a large number of frames are deleted, we can repeat embedding the same secret bits in different segments of frames to solve this problem.
\begin{table}[htbp]
	\centering
	\caption{Extraction Accuracy of The Proposed Algorithm Under The Frame Deletion Attack}
	\begin{tabular}{ccc}
		\toprule
		\multicolumn{1}{p{12em}}{Number of deleted frames} & \multicolumn{1}{c}{Class B} & \multicolumn{1}{c}{Class E} \\
		\midrule
		30    & 89.0\% & 94.0\% \\
		50    & 87.9\% & 89.7\% \\
		\bottomrule
	\end{tabular}%
	\label{tab7}%
\end{table}%

\subsection{Security}
\subsubsection{Objective security}
\par The objective security of the proposed algorithm is discussed in the following two parts. 
\par (1) The security of secret information. Like the coverless image steganography, the proposed coverless video steganography accomplishes the process of information hiding by establishing the mapping relationship between secret information and carriers. Due to no modification of carriers, the proposed algorithm cannot be detected by steganalysis algorithms, which guarantees the security of secret information.
\par (2) The security of auxiliary information. As described before, the auxiliary information needs to be sent to the receivers with the carriers. For security, AES algorithm is adopted to encrypt auxiliary information before transmission, which indicates that in order to obtain secret information, a key shared between the receiver and the sender also needs to be acquired.
\subsubsection{Subjective security}
\par As discussed in Section \uppercase\expandafter{\romannumeral2}, a subjective security test is done to collect people’s views on the transmission of different carriers. And the result indicates that it is securer to utilize a video as a carrier compared with large number of images, which improves the subjective security of transmission. Besides, the video selection rule discussed in Section \uppercase\expandafter{\romannumeral4} further improves the subjective security of transmission, as in this way a smaller number of videos can carry more secret information.

\section{Conclusion}
\par In this paper, a novel coverless video steganography algorithm based on maximum DC coefficients is proposed. The subjective security of coverless steganography is tested for the first time, which shows that the video has more advantages than images. Furthermore, a Gaussian distribution model of DC coefficients considering video coding process is built to analyze the change of DC coefficients before and after video compression. Based on the distribution model, a novel hash sequence generation method is proposed, and the video index structure is established accordingly. After that, the video whose hash sequence equals to the secret bits is chosen as the carrier. Besides, a new practical and fair metric, relative effective capacity is proposed for evaluating the capacity of steganography algorithms, which considers the cost of auxiliary information. Experimental results and analysis show that the proposed algorithm performs better than the state-of-the-art algorithms in terms of capacity, robustness especially to video compression, and security overall. The further work will be focused on the improvement of relative effective capacity and the robustness against more different video attacks such as frame rate changes and transcoding.


%

\ifCLASSOPTIONcaptionsoff
  \newpage
\fi



%

%
%

\bibliographystyle{IEEEtran}
\bibliography{IEEEabrv,mylib}

%

\begin{IEEEbiography}
	[{\includegraphics[width=1in, height=1.25in,clip,keepaspectratio]{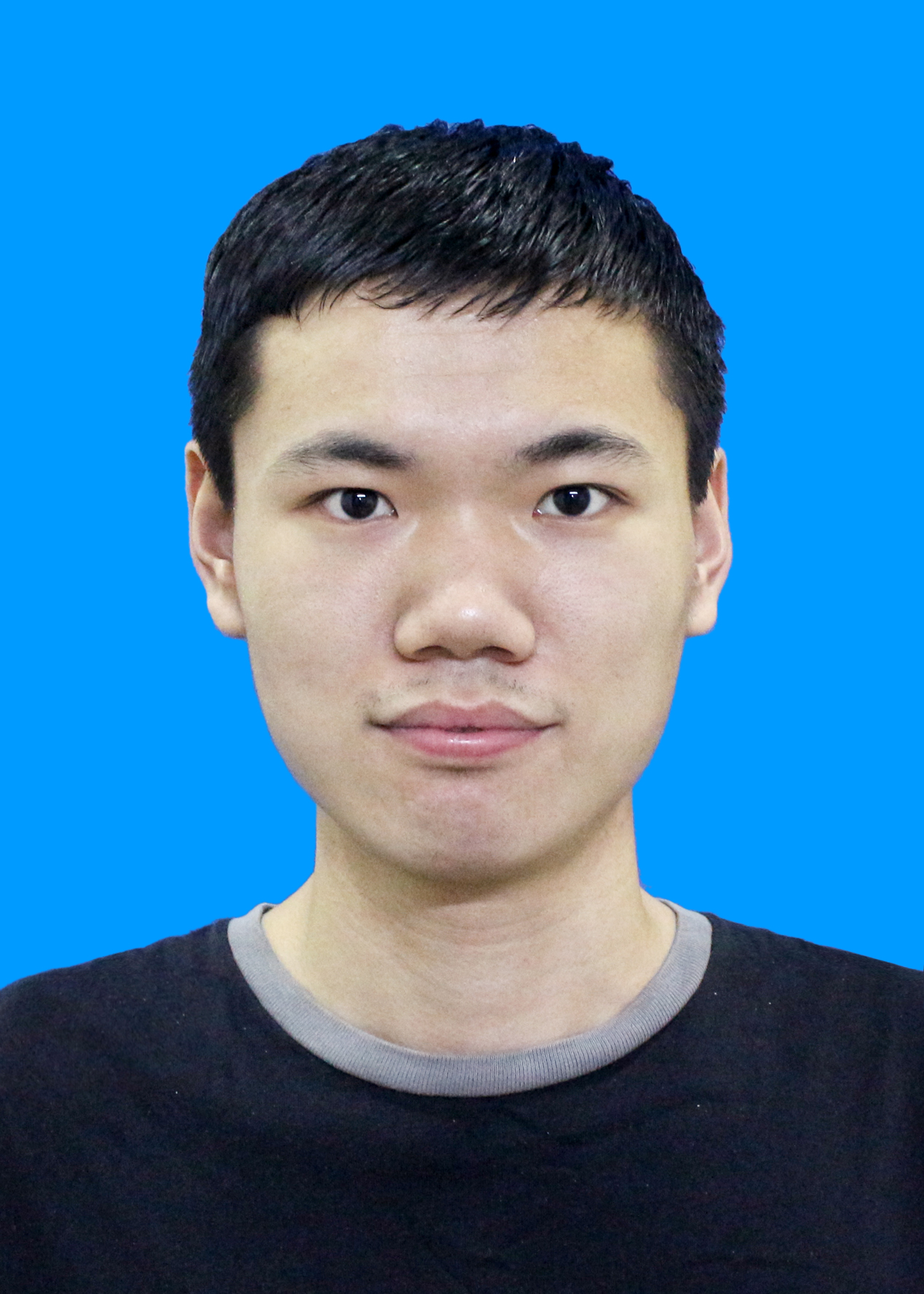}}]{Laijin Meng}
received the B.S. degree in communication engineering from Beijing Jiaotong University, Beijing, China, in 2020.
He is currently pursuing the Ph.D. degree with the School of Electronic Information and Electrical Engineering, Shanghai Jiao Tong University, Shanghai, China. His research interests include steganography and steganalysis.
\end{IEEEbiography}

\begin{IEEEbiography}
	[{\includegraphics[width=1in, height=1.25in,clip,keepaspectratio]{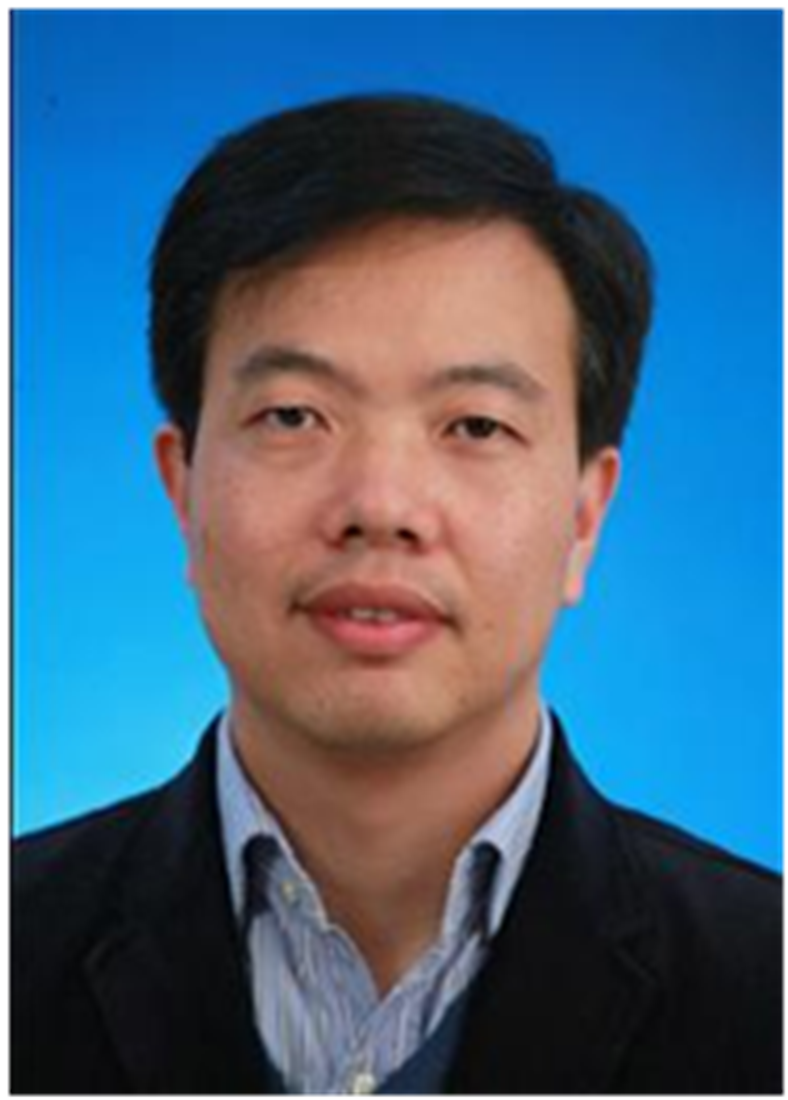}}]{Xinghao Jiang}
	received the Ph.D. degree in electronic science and technology from Zhejiang University, Hangzhou, China, in 2003.
	\par He is currently a Professor with the School of Electronic Information and Electrical Engineering, Shanghai Jiao Tong University, Shanghai, China. He was a visiting scholar with the New Jersey Institute of Technology, Newark, NJ, USA, from 2011 to 2012.  His research interests include multimedia security, intelligent information processing, cyber information security, information hiding and watermarking. Dr. Jiang is an IEEE Senior member.
\end{IEEEbiography}

\begin{IEEEbiography}
	[{\includegraphics[width=1in, height=1.25in,clip,keepaspectratio]{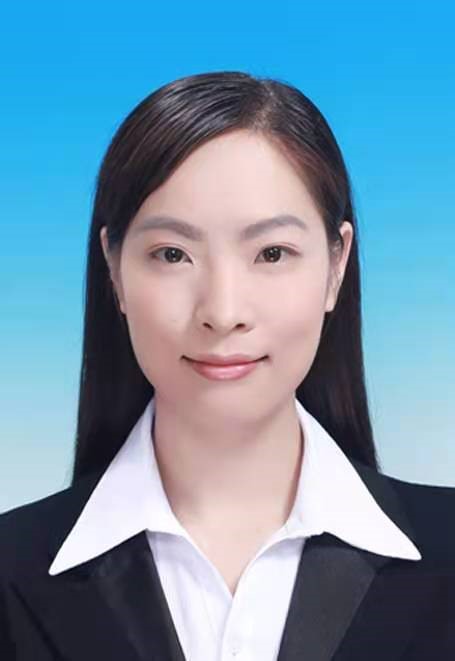}}]{Zhenzhen Zhang}
	received the B.S. degree in Communication Engineering from Zhengzhou University in 2009, and the Ph.D. degree in Circuits and Systems from Beijing Jiaotong University in 2017. After the completion of her Ph.D. program, she has been working as a lecturer in Beijing Institute of Graphic Communication. Her research interests include video steganography and video forensics.
\end{IEEEbiography}

\begin{IEEEbiography}
	[{\includegraphics[width=1in, height=1.25in,clip,keepaspectratio]{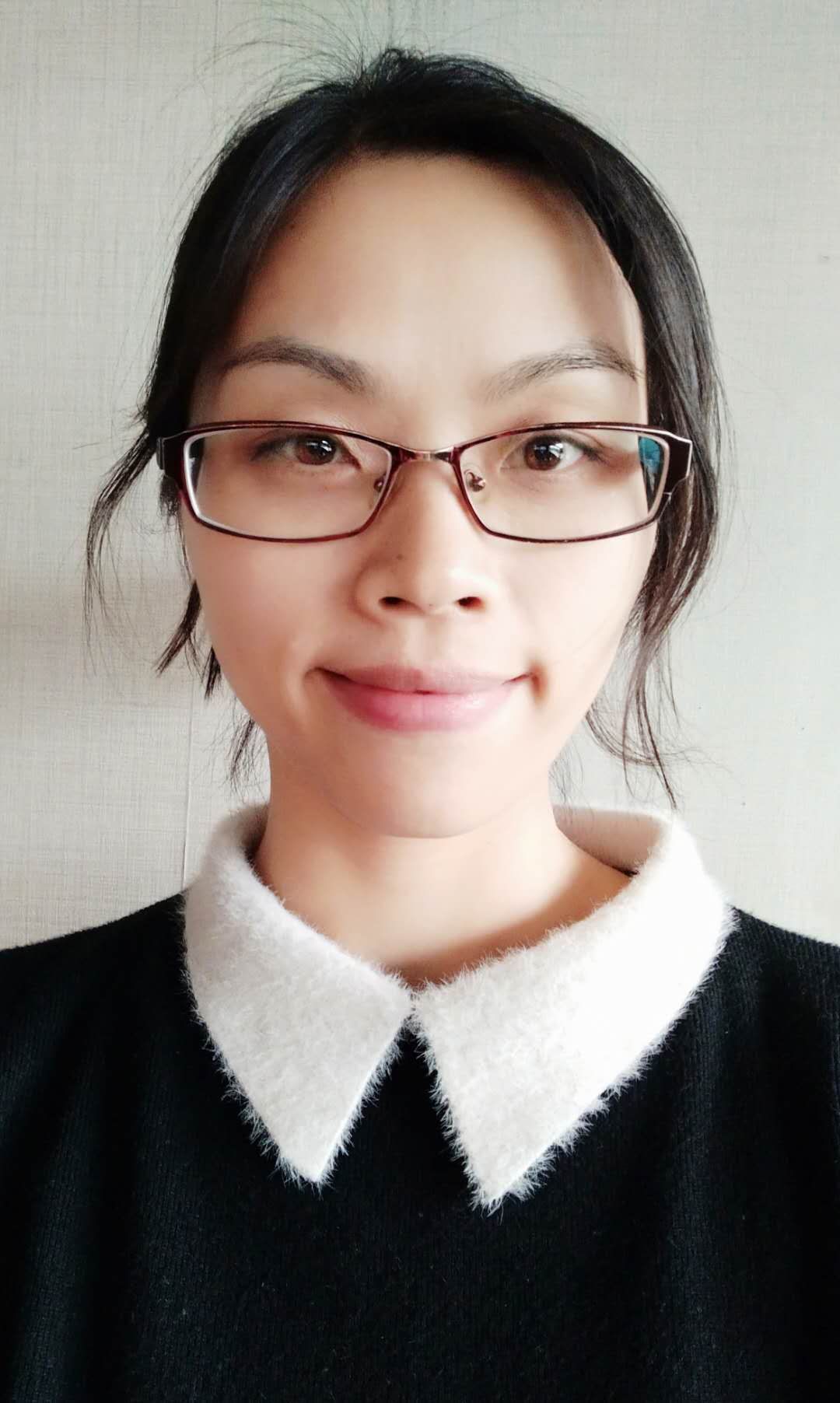}}]{Zhaohong Li}
	received the B.S. degree in communication engineering from Southwest Jiaotong University in 2003 and the Ph.D. degree in traffic information engineering and control from Beijing Jiaotong University in 2008. Since 2008, she has been with the School of Electronic and Information Engineering, Beijing Jiaotong University, where she is currently an Associate Professor. Her research interests include video forensics, information hiding, watermarking, and anti-forgery.
\end{IEEEbiography}

\begin{IEEEbiography}
	[{\includegraphics[width=1in, height=1.25in,clip,keepaspectratio]{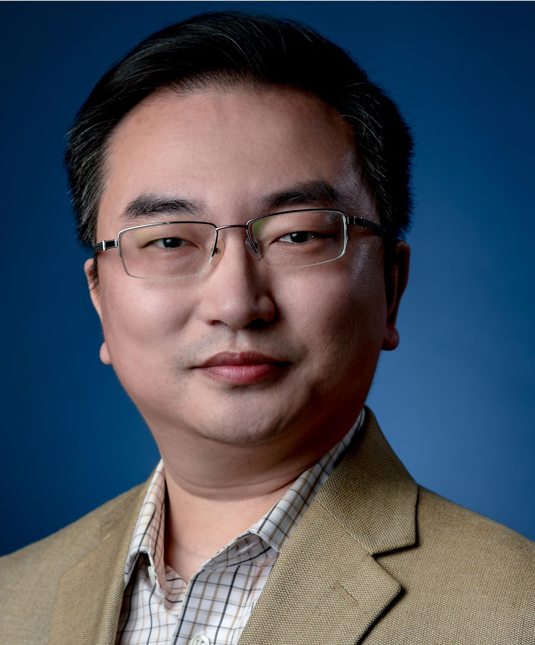}}]{Tanfeng Sun}
	received the Ph.D. degree in information and communication system from Jilin University, Changchun, China, in 2003.
	\par He is currently an Associate Professor with the School of Electronic Information and Electrical Engineering, Shanghai Jiao Tong University, Shanghai, China. He had cooperated with Prof. Y.Q. Shi in New Jersey Institute of Technology, U.S.A, as a visiting scholar from Jul. 2012 to Dec. 2013. His research interests include Digital Forensics on Video Forgery, Digital Video Steganography and Steganalysis, Watermarking, and so on.
	
\end{IEEEbiography}







\end{document}